\documentclass[12pt,a4paper]{article}
\usepackage{jheppub}
\usepackage{graphicx}
\usepackage{lipsum}
\usepackage{slashed}
\usepackage{mathtools}
\usepackage{bm}
\usepackage[all]{hypcap}
\usepackage{tikz}
\usepackage{quiver}
\usepackage{hyperref}
\usepackage{soul}
\usepackage{bbm}
\usepackage{xcolor}

\allowdisplaybreaks

\def\Z{\mathbb{Z}}
\def\R{\mathbb{R}}

\def\Vol{\mathrm{Vol}_{S^2}}

\def\SU{\text{SU}}
\def\SO{\text{SO}}
\def\Spin{\text{Spin}}
\def\U{\mathrm{U}(1)}
\def\su{\mathfrak{su}}

\def\Tr{\mathrm{Tr~}}

\def\ext{\text{Ext}}
\def\sq{\text{~Sq}}

\usepackage{xspace}

\newcommand{\be}{\begin{equation}}
\newcommand{\ee}{\end{equation}}
\newcommand{\bea}{\begin{eqnarray}}
\newcommand{\eea}{\end{eqnarray}}
\newcommand{\beq}{\begin{equation}}
\newcommand{\eeq}{\end{equation}}

\newcommand{\cL}{{\cal L}}

\title{WZW terms without anomalies: \\ generalised symmetries in chiral Lagrangians}

\author[a]{Joe Davighi}
\author[b]{and Nakarin Lohitsiri}

\emailAdd{joseph.davighi@cern.ch}
\emailAdd{nakarin.lohitsiri@durham.ac.uk}

\affiliation[a]{Theoretical Physics Department, CERN, 1211 Geneva 23, Switzerland}
\affiliation[b]{Department of Mathematical Sciences, Durham University,
Upper Mountjoy, Stockton Road, Durham, DH1 3LE, United Kingdom}

\abstract{ 
We consider a 4d non-linear sigma model on the coset $(\SU(N)_L \times \SU(N)_R \times \SU(2))/(\SU(N)_{L+R}\times \U)\cong \SU(N) \times S^2$, that features a topological Wess--Zumino--Witten (WZW) term whose curvature is $\frac{n}{24\pi^2}\mathrm{Tr}(g^{-1}dg)^3 \wedge \mathrm{Vol}_{S^2}$ where $g$ is the $\SU(N)$ pion field. This WZW term, unlike its familiar cousin in QCD, does not match any chiral anomaly,
so its microscopic origin is not obviously QCD-like. We find that generalised symmetries provide a key to unlocking a UV completion.
The $S^2$ winding number bestows the theory with a 1-form symmetry, and the WZW term intertwines this with the $\SU(N)^2$ flavour symmetry into a 2-group global symmetry. Like a 't Hooft anomaly, the 2-group symmetry should match between UV and IR, precluding QCD-like completions that otherwise give the right pion manifold. 
We instead construct a weakly-coupled UV completion that matches the 2-group symmetry, in which an abelian gauge field connects the QCD baryon number current to the winding number current of a $\mathbb{C}P^1$ model, and explicitly show how the mixed WZW term arises upon flowing to the IR. The coefficient is fixed to be the number of QCD colours and, strikingly, this matching must be `tree-level exact' to satisfy a quantization condition.
We discuss generalisations, and elucidate the more intricate generalised symmetry structure that arises upon gauging an anomaly-free subgroup of $\SU(N)_{L+R}$. 
This WZW term may even play a phenomenological role as a portal to a dark sector, that determines the relic abundance of dark matter.
}

\begin{document}

\begin{flushright}
 CERN-TH-2024-120 \end{flushright}

\maketitle

\section{Introduction}

Symmetries provide powerful tools for understanding the structure and dynamics of strongly interacting quantum theories. Beginning in the 1960s, current algebra proved extremely successful in explaining the organising structure of the light pseudo-scalar mesons in quantum chromodynamics (QCD)~\cite{Nambu:1960xd,Gell-Mann:1960mvl}. This was famously refined after the discovery of the chiral anomaly by Adler~\cite{Adler:1969gk}, Bell and Jackiw~\cite{Bell:1969ts}, through which quantum effects were shown to modify the Ward identities one would otherwise infer using the classical Lagrangian. Moving into the 1970s, Wess and Zumino deployed this anomalous current algebra to explain the observed violation of na\"ive spatial parity symmetry and pion number mod 2~\cite{Wess:1971yu}, in {\em e.g.} the decays of the $\phi$ meson to both $K^+K^-$ and $\pi^0 \pi^+ \pi^-$ final states. The anomaly was also shown to account for why the $\eta^\prime$ meson was heavier than its apparent sibling the $\eta$ meson~\cite{tHooft:1976rip,Witten:1979vv,Veneziano:1979ec}. The topological Wess--Zumino--Witten (WZW) term in the action matches this anomalous current algebra relation in the IR, furnishing the (previously local) current algebra structure with a global aspect~\cite{Witten:1983tw}.

Fast forward half a century, and symmetry continues to offer rich and surprising insights into the dynamics of QFTs. A key development in the last decade has been the discovery of {\em generalised} notions of symmetry~\cite{Gaiotto:2014kfa}, beyond the action of groups on local operators. One example is higher $p$-form symmetries, in which abelian groups act not on local operators but on extended objects. From the viewpoint of current algebra, the corresponding symmetry currents for $p$-form symmetries (in the case of continuous symmetry) are not vectors $j^{(1)}_\mu$ (better, 1-forms) as they are for `ordinary' $0$-form symmetries, but tensors $j^{(p+1)}_{\mu...\sigma}$ with more than one index (specifically, $(p+1)$-forms). Returning to QCD and its discrete symmetries, $\SU(n)$ gauge theory possesses such a $\mathbb{Z}_n$ valued 1-form symmetry, which was shown by Gaiotto, Kapustin, Komargodski and Seiberg to have a mixed anomaly with parity when the QCD theta angle equals $\pi$~\cite{Gaiotto:2017yup}, for $n$ even. This subtle anomaly in QCD, discovered nearly 50 years after the ABJ anomaly, means that QCD at $\theta = \pi$ cannot be trivial in the deep infrared (IR). 

It has also been recently understood that there can be a non-trivial current algebra between higher-form symmetries of different degree. The simplest example is known as a 2-group symmetry structure, wherein 0-form and 1-form symmetries `mix', as first studied in topological field theory by Kapustin and Thorngren~\cite{Kapustin:2013uxa}, based on the idea of higher gauge theory \cite{Baez:2004in, Baez:2005qu, Baez:2010ya}. This was further elucidated in the wider context of generalised symmetries by Sharpe~\cite{Sharpe:2015mja}. In the case where both 0-form and 1-form symmetries are continuous, this can be captured, following the tradition of current algebra, by a Ward identity of the form~\cite{Cordova:2020tij}:
\begin{equation} \label{eq:intro_algebra}
    \langle i\partial^\mu j^{(1) a}_\mu(x) j^{(1) b}_\nu(y) \rangle = \langle -\delta(x-y) f^{abc} j^{(1)c}_\nu +  \frac{n}{8\pi^2}\delta^{ab}\partial^\mu \delta(x-y) j^{(2)}_{\mu\nu}(y) \rangle\, , \quad n \in \Z\, ,
\end{equation}
where $f^{abc}$ are the structure constants for the 0-form flavour symmetry. The fact that the coefficient $n$ is an integer is reminiscent of the original anomalous current algebra relation. (Here $n\in \Z$ really labels the `Postnikov class' characterizing the particular 2-group, and is an element in an appropriate cohomology group.)
Indeed, like the anomaly, it implies that such a 2-group structure should be preserved along the RG flow, or else the symmetry be broken.

This paper concerns a seemingly innocuous extension of 4d QCD (let's say, for concreteness, that the UV theory contains an $\SU(n_c)$ gauge group and $N$ fundamental Dirac fermions)
by a pair of extra pions in the infrared that live on $S^2$. This theory admits a second topological term in the low-energy effective action that involves both QCD pions and the extra pions on $S^2$, as recently observed in Ref.~\cite{Davighi:2024zip}.
Like the original WZW term, this topological term has an integer-quantized coefficient. But unlike the WZW term, this term does not match any chiral anomaly -- so it has no obvious interpretation as originating from a loop of chiral fermions in the microscopic theory. 

In this paper, we show that this `mixed WZW term' actually encodes a 2-group global symmetry structure in the IR theory, which mixes the QCD flavour symmetry with a 1-form symmetry associated to the winding number around the $S^2$ factor of the target space. The coefficient $n$ of this mixed WZW term is precisely the Postnikov class encoding the 2-group symmetry. This furnishes an interesting new example of continuous 2-group symmetry in a 4d quantum field theory, that contains only scalar degrees of freedom in the infrared.\footnote{Other examples of continuous 2-group symmetry without fermions include continuous 2-groups in hydrodynamics and holography \cite{Iqbal:2020lrt, DeWolfe:2020uzb}.}

The observation of a non-trivial 2-group global symmetry in this theory puts strong constraints on possible RG flows that can realise this phase in the IR.\footnote{The rigidity of higher-group symmetry structures mean they can also be tracked across dualities, as well as RG flows, to provide highly non-trivial checks -- see for instance~\cite{Lee:2021crt}. } 
In particular, a UV completion that exactly preserves the QCD flavour symmetry must, in order to close the 2-group current algebra relation (\ref{eq:intro_algebra}), at the very least possess a continuous 1-form symmetry. At a stroke, this precludes QCD-like UV completions that one might guess, such as a confining $\SU \times \SO$ gauge theory or an $\SU$ gauge theory with both fundamental and adjoint quarks condensing. 

We then propose a UV completion of this mixed WZW term, consistent with the 2-group symmetry constraints, in which an abelian $\U_g$ gauge field that is Higgsed in the IR connects the QCD sector to the $S^2$ sector. 
The scalars on $S^2$ are embedded in a linear sigma model of two complex scalars at high energy, charged equally under $\U_g$, which condense to Higgs the $\U_g$ and simultaneously break a global $\SU(2)$ symmetry down to a $\U$ subgroup, delivering the massless pions on $\SU(2)/\U\cong S^2$. 
Once the abelian gauge field is `integrated in' as we go to higher energies, the $S^2$ winding number is precisely traded for the abelian monopole flux. 
On the quark side, the $\U_g$ gauge field couples to baryon number, which becomes identified with the topologically conserved Skyrme current~\cite{Skyrme:1961vq,Balachandran:1982dw,Witten:1983tx} in the IR. These very particular couplings of the $\U_g$ gauge field are the ingredients needed to match onto the mixed WZW topological term, simply by integrating out a weakly coupled gauge field at tree-level. Notably, because the coefficient of the mixed WZW term is quantized for consistency of the low-energy effective action (as mentioned, it is a class in integral cohomology), this tree-level matching formula cannot receive loop correction. In other words, the matching must be tree-level exact.

When an anomaly-free abelian subgroup of the QCD flavour symmetry is also gauged (such as gauging QED in real-world QCD), the symmetry structure becomes more intricate. We show how additional non-invertible symmetries arise (beyond those corresponding to the `usual' ABJ anomaly in pure QCD), while a remnant of the original 2-group global symmetry remains.

Finally, we remark that the identification of this peculiar mixed WZW term in Ref.~\cite{Davighi:2024zip} was motivated by the fact that it can play an important role in phenomenology. It was there shown that the extra pions on $S^2$ could constitute the dark matter (DM) in our Universe, with the mixed WZW term providing an (almost) unique portal from this dark sector to low-energy QCD that is topological.\footnote{Other WZW-like topological portals between low-energy QCD + electromagnetism and a dark sector could (a) connect QCD baryon number to a dark photon, or (b) connect the visible photon to a dark baryon number current. There could be yet further topological portals that are, however, more like $\theta$-terms, for instance mixing the visible photon with a dark photon; but these would be total derivatives and so not give rise to local interactions between the visible and dark particles. }
This portal can reproduce the observed abundance of dark matter today via thermal freeze-out.\footnote{The topological nature of the interaction is not just a theoretical nicety, but plays a crucial role in the phenomenology: being anti-symmetrized in field indices completely suppresses elastic interaction channels that would otherwise lead to strong constraints from DM direct and indirect detection. This allows the off-diagonal `co-annihilation' channel to dominate, realising the light thermal inelastic DM scenario~\cite{Tucker-Smith:2001myb}. }
The present paper offers one path to UV completing this novel portal, guided by the identification of a rich generalised symmetry structure, that will be crucial to understanding the full phenomenological implications (both in cosmology and collider experiments) of this topological portal EFT.

The rest of the paper is as follows. In \S \ref{sec:WZW} we recall the construction of the mixed WZW term in the low-energy EFT of QCD extended by pions on $S^2$, and we identify the presence of 2-group symmetry encoded by this term. We then use this symmetry structure to investigate the UV completion of this EFT: in \S \ref{sec:no-go} we discuss QCD-like `non-completions', before setting out the weakly coupled UV completion in \S \ref{sec:proper-completion} wherein QCD is coupled to a linear sigma model by an abelian gauge field. We consider variations of the scalar sector, in particular a generalisation of our story obtained by replacing the $S^2$ factor by a general complex projective space, in \S \ref{sec:scalar_variation}. Finally, in \S \ref{sec:gauging} we treat the gauged case, before concluding.

\section{Mixed WZW term on \texorpdfstring{$\mathrm{SU}(N)\times S^2$}{SU(N)×S²}} \label{sec:WZW}

Our story starts with a low-energy effective field theory (EFT) of pions on the manifold $\SU(N) \times S^2$, in $3+1$ dimensions. This would arise from an ultraviolet theory with approximate global symmetry of the product form
$G = \SU(N)_L \times \SU(N)_R \times \SU(2)_D$,
which is spontaneously broken down to
    $H = \SU(N)_{L+R} \times \U_D$
as we flow to the infrared, where $\SU(N)_{L+R}$ is the diagonal subgroup of $\SU(N)_L \times \SU(N)_R$ and $\U_D\subset \SU(2)_D$.

Classifications of Wess--Zumino--Witten (WZW) terms~\cite{Wess:1971yu,Witten:1983tw} via generalized cohomology~\cite{Freed:2006mx,Davighi:2018inx,Lee:2020ojw,Yonekura:2020upo} tell us there is a {\em mixed} WZW interaction coming from the existence of a $G$-invariant closed 5-form involving pions on both the $\SU(N)$ and $S^2$ factors, namely~\cite{Davighi:2024zip}
\begin{equation} \label{eq:omega}
    \omega = n\omega_3 \wedge \mathrm{Vol}_{S^2}, \qquad \omega_3 = \frac{1}{24\pi^2} \mathrm{Tr}(g^{-1}dg)^3\, , \qquad n \in \Z\, ,
\end{equation}
where $\Vol$ is the volume form on the $S^2$. To define this term it is easier to start with a homological description -- we briefly discuss the refinement via bordism afterwards.
Given there are no homologically non-trivial 4-cycles in the target space, since $H_4(\SU(N) \times S^2)=0$,
any 4-cycle in the target space obtained by pushing forward $\Sigma_4$ (more precisely, pushing forward a cycle in the fundamental class $[\Sigma_4]$) can be realised as the boundary of a 5-cochain $X_5$, on which we can integrate $\omega$ to obtain the exponentiated action \`a la Witten~\cite{Witten:1983tw}
\begin{align} \label{eq:witten}
\exp\left(iS[\Sigma_4=\partial X_5]\right) 
&=\exp\left(2\pi i\int_{X_5} \omega\right) \\
&= \exp\left(2\pi i\int_{X_5} \frac{n}{24\pi^2} \mathrm{Tr}(g^{-1}dg)^3 \wedge \Vol \right)\, . \nonumber
\end{align}
The normalisation is such that $\omega$ is an integral 5-form on $\SU(N) \times S^2$, meaning that $e^{2\pi i\int_{z_5}\omega}=1$ for any 5-cycle $z_5$, which guarantees the exponentiated action defined in this way is independent of the choice of `bulk manifold' (more precisely, 5-cochain) $X_5$~\cite{Witten:1983tw}. 

A local expression for the 4d Lagrangian can be obtained by 
expanding the QCD pion field locally as 
\begin{equation}
g(x)=\exp(2i\pi_a(x)t^a/f_\pi) = 1+ 2i\pi_a (x) t^a/f_\pi + \dots\, ,    
\end{equation}
and taking $\chi_i/f_D$ as local Cartesian coordinates in the vicinity of a given vacuum point ($\chi_i=0$) on the $S^2$ factor. Then the $\SU(N)$-invariant 3-form is expanded as
\begin{equation}
    \omega_3 = \frac{1}{24\pi^2}\frac{2}{f_\pi^3}f_{abc}d\pi_a \wedge d\pi_b \wedge d\pi_c + \mathcal{O}(\pi^4)\, ,
\end{equation}
where $f_{abc}$ are the $\SU(N)$ structure constants,
while the volume form on the $S^2$ is expressed as 
\begin{equation}
    \Vol = \frac{1}{4\pi}\frac{1}{f_D^2} \cos(\chi_1) d\chi_1 \wedge d\chi_2 = \frac{1}{4\pi} \frac{1}{f_D^2}  d\chi_1 \wedge d\chi_2 +\mathcal{O}(\chi^3) \, .
\end{equation}
On the coordinate patch near the origin $(\pi_a,\chi_i)=0$, we can use Stokes' theorem to get a local expression for the Lagrangian (also including a factor $2\pi i$):
\begin{equation}
    \cL =  \frac{in\epsilon^{\mu\nu\rho\sigma}}{48\pi^2 f_D^2 f_\pi^3} f_{abc} \epsilon_{ij} \pi_a \partial_\mu \pi_b \partial_\nu \pi_c \partial_\rho \chi_i \partial_\sigma \chi_j + \mathcal{O}(\pi^4 \chi^2, \pi^3 \chi^3)\, ,
    \label{eq:pert-mixed-WZW}
\end{equation}
which was used to calculate scattering cross-sections between the $\pi$ and $\chi$ pions in~\cite{Davighi:2024zip}.

\subsection{Remarks on bordism {\em vs} homology, and a discrete theta angle}

To be more precise, one should re-formulate the above construction of a WZW-like action, which involves realising spacetime $\Sigma$ as a boundary of a bulk in one higher dimension, more directly using the language of bordism. This is somewhat tangential to the main story of this paper, so we limit ourselves to some cursory remarks.

Roughly speaking, bordism tells us whether manifolds equipped with certain structures --- gauge bundles, spin structures, metrics --- can be written as boundaries of manifolds in one dimension higher, with all structures smoothly extended thereto. In short, bordism deals directly with manifolds themselves, rather than passing to cochains as in the homological description above. The UV theory we have in mind will feature fermions, so the appropriate bordism theory is (reduced) spin-bordism. 

In Appendix~\ref{app:bordism-comp} we use the Adams spectral sequence~\cite{Adams:1958} to compute
that the relevant reduced bordism group for us here does not vanish, but is given by 
\begin{equation} \label{eq:Om4}
    \tilde{\Omega}_4^{\mathrm{Spin}}\left(\SU(N) \times S^2\right) \cong \tilde{\Omega}_4^{\mathrm{Spin}}\left(S^2\right) \cong \Z_2\, .
\end{equation}
This means there is a class of spin 4-manifolds $[X_4]$, together with maps $\sigma$ to $\SU(N)\times S^2$, that cannot be extended to spin 5-manifolds also equipped with such maps. This equivalence class is the generator of the bordism group \eqref{eq:Om4}. The existence of manifolds that are not boundaries is an obstruction to a Witten-like definition \eqref{eq:witten} of the WZW term when evaluated on such manifolds. This obstruction is not seen using homology.

We can nonetheless proceed in attempting to define our WZW term on a general manifold as follows.
That the bordism group is $\Z_2$ means that the union of two generators is a boundary of some closed 5-manifold $X_5$ with the map $\sigma$ extended. The exponentiated action evaluated on this union $X_4 \sqcup X_4$ must be the square of the action evaluated on the original $X_4$, by locality. Therefore, we can define the exponentiated WZW action for this union as
\begin{equation} \label{eq:X4squared}
    \left(\exp(i S[X_4])\right)^2 = \exp \left( 2\pi i \int_{X_5} \omega\right)
\end{equation}
Taking the square root of the right-hand-side then gives two branches of solution, and we must make a consistent choice of branch to define the theory on all possible 4-manifolds $X_4$ equipped with background structures. This choice, which is equivalent to specifying a $\Z_2$-valued parameter of the theory (also known as a `discrete theta angle'), can be made by fixing a representative $X_4^{(0)}$ for the non-trivial class in $\tilde{\Omega}_4^{\mathrm{Spin}}\left(\SU(N) \times S^2\right) \cong \Z_2$ and choosing a branch for $\exp(iS[X^{(0)}_4])$. For instance, we can fix $X_4^{(0)}$ to be a generator of $\tilde{\Omega}_4^{\mathrm{Spin}}\left(S^2\right) \cong \Z_2$ with the background $\SU(N)$ pion configuration set to the trivial map {\em i.e.} $\pi_a(x)=0$, in which case the RHS of \eqref{eq:X4squared} evaluates to unity. Then we can choose either
\begin{equation} \label{eq:sign}
\exp(iS[X_4^{(0)}])=\pm 1\, .
\end{equation}
For other manifolds in the same bordism class, for which the topological term in general evaluates to a $\mathbb{C}$-number, the mixed WZW can be defined using a bordism between $X^{(0)}_4$ and $X_4$, and integrating the differential form $\omega$ thereon.

The choice of sign in \eqref{eq:sign} is in fact fixed by non-perturbative 't Hooft anomaly matching: the solution with a minus sign must be chosen if the symmetry group $\SU(2)$ suffers from the mod 2 global anomaly of Witten~\cite{Witten:1982fp} in the UV. 
To see this, let us be more explicit in our choice of representative and take $X_4^{(0)}$ to be $S^4$, equipped with a homotopically non-trivial dark pion map $U_\chi:S^4 \to \SU(2)/\U \cong S^2$, $[U_\chi]=1 \mod 2 \in \pi_4(S^2) \cong \Z_2$. The negative sign choice in \eqref{eq:sign}, which is the non-trivial value of this discrete theta angle, corresponds to assigning a partition function phase $(-1)^{[U_\chi]}$ that depends on the homotopy class of the dark pion map, while the positive sign choice assigns the trivial phase to all maps. Now let us discuss how this matches the Witten anomaly.
Under a local $\SU(2)$ transformation specified by map $g(x):S^4 \to \SU(2) \cong S^3$, the pion field transforms under the left-action $U_\chi \to g \cdot U_\chi$. We know that, if the UV theory has non-zero Witten anomaly, then the partition function should flip sign under background gauge transformations for which $[g]=1\mod 2 \in \pi_4(S^3)$, for any value of the background gauge field (including zero). Our negative-sign discrete theta angle has this property, as follows. Using the Hopf fibration $f:S^3 \to S^2$, we can pullback $U_\chi$ to a map $f^\ast U_\chi:S^4 \to S^3$, and the associated long exact sequence in homotopy tells us that $[f^\ast U_\chi]$ is non-trivial in $\pi_4(S^3)$ iff $U_\chi$ is non-trivial in $\pi_4(S^2)$. Thus we can write our discrete WZW term as $(-1)^{[f^\ast U_\chi]}$. Under $g(x)$, this is sent to $(-1)^{[g \cdot (f^\ast U_\chi)]} = (-1)^{[g] \cdot [f^\ast U_\chi]} = (-1)^{[g]}(-1)^{[f^\ast U_\chi]}$, thus the partition function flips sign, matching the anomaly. On the other hand for the trivial value of the discrete theta angle, {\em i.e.} the positive sign choice, one simply assigns the trivial phase to all paths and so there is no anomalous variation. For more detail on matching global anomalies with discrete WZW terms, we refer the reader to Refs. \cite{Yonekura:2020upo,Lee:2020ojw}.

In principle there is also a subtlety concerning the normalisation of the WZW term, related to whether one uses a classification based on homology, bordism, or even homotopy groups.\footnote{For example, the normalisation of the usual WZW term of 4d QCD is subject to a factor of two difference~\cite{bott1958space} if one chooses to normalise the curvature form $\propto \Tr (g^{-1} dg)^5$ against the generator of the homotopy group $\pi_5(\SU(N))=\Z$ or against the integral homology $H_5(\SU(N);\Z)=\Z$; furthermore, the homotopy-based normalisation happens to agree with that determined by spin-bordism~\cite{Freed:2006mx,Lee:2020ojw}, which is arguably the most justifiable choice. }
In our case, the normalisation encoded in our formulae \eqref{eq:omega} above is that prescribed by homology; nonetheless, later we show that this mixed WZW term relates to an anomaly polynomial~\eqref{eq:Phi6} for a particular fermionic theory, whose normalisation condition is determined by the Atiyah--Singer index theorem~\cite{atiyah1968indexI,atiyah1968indexII,atiyah1968indexIII}. This verifies that the normalisation of \eqref{eq:omega} coincides with the correct normalisation in spin bordism.

\subsection{An anomaly-matching puzzle} \label{sec:puzzle}

We now continue with the main line of argument, and consider the symmetry/anomaly information encoded in this mixed WZW term.
Recall the coefficient of the ordinary WZW term for pure QCD, associated with the 5-form $\omega_5\propto \mathrm{Tr}(g^{-1}dg)^5$, is fixed by 't Hooft anomaly matching~\cite{tHooft:1979rat} for the anomalous chiral symmetry currents generating $\SU(N)_L$ and $\SU(N)_R$ separately. The UV anomalies come from terms in the anomaly polynomial $\Phi_6 \sim \mathrm{Tr}(F_L^3) + \mathrm{Tr}(F_R^3)$. When QED is gauged, this translates into an ABJ anomaly that explicitly breaks the chiral flavour symmetries.

We can similarly investigate the possible anomaly-matching role of the mixed WZW term by turning on gauge currents for various global symmetries.
For instance, upon gauging QED one also gets the term
\begin{equation}
    \cL \sim \frac{1}{f_\pi f_D^2}\epsilon^{\mu\nu\rho\sigma} \pi_0 F_{\mu\nu} \partial_\rho \chi_1 \partial_\sigma \chi_2\, ,
\end{equation}
which was put to phenomenological use as a possible portal to dark matter in~\cite{Davighi:2024zip}.
If one also gauged the dark $\U_D$ unbroken global symmetry, 
one would further get a term
\begin{equation}
    \cL \sim \frac{1}{f_\pi}\epsilon^{\mu\nu\rho\sigma} \pi_0 F_{\mu\nu} (F_D)_{\rho\sigma}\, .
\end{equation}
Thus, the mixed WZW term na\"ively matches mixed anomalies corresponding to anomaly polynomial terms
    $\Phi_6 \sim a_L\mathrm{Tr}(F_L^2 F_D) + a_R\mathrm{Tr}(F_R^2 F_D)$.
One might try to conclude that this EFT term arises in UV theories that feature a mixed anomaly between the `light' and `dark' flavour symmetries. But these mixed anomaly coefficients $a_{L/R}$ {\em vanish} if the $\U_D$ representations are embedded in $\SU(2)_D$, because there is of course no mixed anomaly between $\SU(2)$ and $\SU(N\geq 3)$ in 4d.
The mixed WZW term in the low-energy EFT is consistent with an exact $\SU(2)_D$ flavour symmetry, and so there should be a UV account of its origin that does not invoke explicit $\SU(2)_D$-breaking as a necessary ingredient. 

The main question we try to answer in this paper is: what is the microscopic origin, if not an anomaly, of the mixed WZW term permitted by this particular EFT?

\subsection{Infrared 2-group symmetry}

Generalised symmetries (in place of anomalies) present a key to unlocking this apparent puzzle. It turns out that the mixed WZW signals the presence of a generalised symmetry structure known as a {\em 2-group symmetry}, which mixes the QCD flavour 0-form symmetry with a 1-form global symmetry in a non-trivial way. 

First, we observe that the volume form on $S^2$ provides us with a topologically conserved ({\em i.e.} closed) non-trivial 2-form, which may be identified with the current for a 1-form winding number symmetry
\begin{align}
    j^{(2)}_{\mathrm{wind}} :&= \star_4 \mathrm{Vol}_{S^2} =  \frac{\sqrt{|\mathrm{det}(g)|}}{4\pi}\epsilon_{\mu\nu\rho\sigma} \cos(\chi_1)\partial^\mu\chi_1 \partial^\nu\chi_2 dx^\rho dx^\sigma , \\
    d\star_4 j_{\text{wind}}^{(2)} &= d\, \mathrm{Vol}_{S^2} = 0\, . \nonumber
\end{align}
Here, $\star_4$ is the Hodge dual on spacetime $(\Sigma,g)$, and indices have been raised using the inverse metric on $\Sigma$, {\em viz.} $\partial^\mu \chi_i=g^{\mu\lambda} \partial_\lambda \chi_i$. 

To understand the global symmetry structure precisely, it is instructive to turn on background gauge fields. Let us start with the 1-form symmetry.
As for a regular 0-form symmetry, one can minimally couple the 1-form symmetry to a background gauge field, which in this case is a 2-form gauge field $B$. The minimal coupling term is
\begin{equation} \label{eq:Bcoup}
    S_{\text{coup}} = 
    i\int_{\Sigma_4} \star_4\, j_{\text{wind}}^{(2)} \wedge B= 
    i\int_{\Sigma_4} \Vol \wedge B
\end{equation}
By construction, the coupling term is invariant under a $\U^{[1]}_{\text{wind}}$ 1-form gauge transformation
\begin{equation}
    B \mapsto B + d \Lambda^{(1)},
\end{equation}
because $\Vol$ is closed.

Now for the 0-form symmetry, we turn on background 1-form gauge fields $A_L$ and $A_R$. The gauging of these 0-form symmetries in the mixed WZW term can be deduced from the gauging of the 3d WZW term $S[\Sigma_2]:=\int_{Y_3} \Tr (g^{-1}dg)^3/24\pi^2$, where $\partial Y_3=\Sigma_2$ which we know yields the 3d Chern--Simons (CS) theory
$\int_{Y_3} \frac{1}{8\pi^2} \left[\mathrm{CS}(A_L) -\mathrm{CS}(A_R)\right]$ 
where $\mathrm{CS}(A)= AdA + \frac{2}{3} A^3$.
For our 4d theory, gauging therefore gives the action 
\begin{equation}
    iS[\Sigma_4=\partial X_5] = \int_{X_5} \frac{-in}{8\pi^2}\left[\mathrm{CS}(A_L) -\mathrm{CS}(A_R)\right] \wedge \Vol
\end{equation}
Crucially, after coupling to the background gauge fields $A_L$, $A_R$, $B$ for $\SU(N)_L$, $\SU(N)_R$, and $\U^{[1]}_{\text{wind}}$, respectively, the mixed WZW term is not gauge invariant.
Rather, under 
the flavour 0-form gauge transformation
\begin{equation} \label{eq:0form-gt}
    A_L \mapsto A_L + D_{A_L} \lambda^{(0)}_L, \qquad  A_R \mapsto A_R + D_{A_R} \lambda^{(0)}_R\,,
\end{equation}
it shifts by 
\begin{equation}
    \delta (2\pi i S) = -\frac{i n}{4\pi}\int_{\Sigma_4} \Vol \wedge \left[\Tr(\lambda^{(0)}_L d A_L)-\Tr(\lambda^{(0)}_R d A_R)\right]\, .
    \label{eq:mixed-WZW-gt}
\end{equation}
One might interpret this as an extra anomalous variation under QCD flavour symmetry, in addition to the anomalous variation already encoded in the usual WZW term of pure QCD. But an important difference compared to the usual WZW term variation is that, for this term, the factor $\Vol$ on the RHS of the anomalous variation is an operator in the theory that does not vanish upon turning off the 0-form background gauge field.

However, now consider modifying the transformation law for the 2-form gauge field to depend on the QCD flavour 0-form transformation, in the following way:
\begin{equation}
    B \mapsto B + d \Lambda^{(1)} + \frac{\hat\kappa_L}{4\pi}\,\Tr \left(\lambda^{(0)}_L dA_L\right) + \frac{\hat\kappa_R}{4\pi}\,\Tr \left(\lambda^{(0)}_R dA_R\right)\, .
\end{equation}
Mathematically, this (together with~\ref{eq:0form-gt}) defines the infinitesimal version of a continuous 2-group gauge transformation,\footnote{We will not actually define what a 2-group is in this paper. For an introduction to the mathematical notion of a 2-group, and how these structures appear in QFT, see {\em e.g.} \S 2 of~\cite{Davighi:2023luh}. }
with the locally-defined differential forms $(A_L, A_R, B)$ forming a 2-connection.
Given the 0-form flavour symmetry $\SU(N)^2$ and the 1-form winding number symmetry valued in $\U^{[1]}$, possible 2-group structures intertwining the two are classified (subject to certain simplifying assumptions that apply here) by a topological invariant called the {\em Postnikov class}, which is the pair
\begin{equation}
    (\hat\kappa_L, \hat\kappa_R) \in H^4(B\SU(N)^2;\Z) \cong \Z \times \Z\ .
\end{equation}
If we postulate this generalised gauge transformation law for the background fields, then the minimal coupling term (\ref{eq:Bcoup}) shifts by
\begin{equation}
\frac{i}{4\pi}\int_{\Sigma_4} \Vol \wedge \left[\hat{\kappa}_L\Tr (\lambda^{(0)}_L dA_L) + \hat{\kappa}_R\Tr (\lambda^{(0)}_R dA_R)\right].
\end{equation}
This is precisely how the mixed WZW term with the $SU(N)_{L/R}$ background gauge fields turned on shifts, in the opposition direction, under the flavour gauge transformation, provided that we identify 
\begin{equation} \label{eq:Post_IR}
    \hat{\kappa}_L = - \hat\kappa_R=n\, .
\end{equation}
The whole combination is then gauge invariant. This is a sign that there is a non-trivial 2-group structure present \cite{Cordova:2018cvg}, and that the coefficient of the mixed WZW term exactly determines the Postnikov class characterizing the `twisting' of this 2-group symmetry.

\subsection*{Ward identities for the 2-group symmetry}

To see that this is not just an artifact of turning on background fields, we can show how this 2-group structure is manifest at the level of current algebra.
We start by deriving the modified conservation laws for the 1-form currents when only the background
fields for the flavour symmetries are turned on. Recall that the currents $j^{(1)}_L$ and $j^{(1)}_R$
for the flavour symmetries $\SU(N)_L$ and $\SU(N)_R$ can be  minimally coupled to their corresponding background gauge fields as 
\begin{equation}
    S_{\text{currents}} = \int_{\Sigma_4} (\Tr \star j^{(1)}_L \wedge A_L  + \Tr \star j^{(1)}_R \wedge A_R)\,.
\end{equation}
Including also the mixed WZW term, and using the gauge transformation of the mixed WZW term given in Eq. \eqref{eq:mixed-WZW-gt}, the total action shifts by
\begin{align}
    \delta S = \int_{\Sigma_4}&\bigg[ \left(D_{A_L} \star j^{(1)}_L - \frac{n}{8\pi^2} \star j^{(2)}_{\text{wind}} \wedge d A_L\right)^a \lambda^a_L \bigg. \\
    &\bigg. \,\, + \left(D_{A_R} \star j^{(1)}_R + \frac{n}{8\pi^2} \star j^{(2)}_{\text{wind}} \wedge d A_R\right)^a \lambda^a_R  \bigg], \nonumber
\end{align}
upon integrating by parts the terms involving 1-form currents. Imposing gauge invariance, we obtain a pair of modified conservation laws:
\begin{align}
    D_{A_L} \star j^{(1)}_L &= \frac{n}{8\pi^2} \star j^{(2)}_{\text{wind}} \wedge d A_L\,,\\
    D_{A_R} \star j^{(1)}_R &= - \frac{n}{8\pi^2} \star j^{(2)}_{\text{wind}} \wedge d A_R\,,
\end{align}
which hold inside the path integral. This means
\begin{equation}
    \int \mathcal{D}\Phi\, \left(D_{A_L} \star j^{(1)}_L - \frac{n}{8\pi^2} \star j^{(2)}_{\text{wind}} \wedge d A_L \right) e^{iS_{\text{current}}} e^{iS_0} = 0,
\end{equation}
where $S_0$ is the original action when all background fields are turned off, and $\Phi$ stands for all the dynamical fields collectively. A similar expression holds for $j^{(1)}_R$.

Current algebra relations for $j^{(1)}_L$ and $j^{(1)}_R$ can then be obtained by taking functional derivatives of these conservation laws with respect to the background fields. Expanding the integrand of the path integral 
above for infinitesimal $A_L$ (with $A_R=0$), we obtain, at linear order,
\begin{equation}
    i \partial_\mu j_L^{(1)a \mu}(x) \int d^4y A^b_{L \nu}(y) j_L^{(1)b\nu}(y) + f^{abc}A^b_L(x) j_L^{c(1)\nu}(x) =\frac{n}{8\pi^2} j^{(2)\lambda\nu}_{\text{wind}}(x) \partial_\lambda A^a_{L\nu} (x)
\end{equation}
Taking the functional derivative of this equation with respect to $A^b_{L\nu}(y)$ gives 
\begin{equation} \label{eq:2grp-algebra-L}
    i\partial_\mu j_L^{(1)a\mu}(x) j_L^{(1)b\nu}(y) + f^{abc} \delta(x-y)j_L^{(1)c\nu}(y) = \frac{n}{8\pi^2} \delta^{ab}\frac{\partial}{\partial x^\lambda} \delta(x-y) j^{(2)\lambda\nu}_{\text{wind}}(y)\,.
\end{equation}
 Repeating the operation with $A_R$, we similarly obtain
\begin{equation} \label{eq:2grp-algebra-R}
    i\partial_\mu j_R^{(1)a\mu}(x) j_R^{(1)b\nu}(y) + f^{abc} \delta(x-y)j_R^{(1)c\nu}(y) = -\frac{n}{8\pi^2} \delta^{ab}\frac{\partial}{\partial x^\lambda} \delta(x-y) j^{(2)\lambda\nu}_{\text{wind}}(y)\,.
\end{equation}
Sure enough, these expressions exactly reproduce the current algebras for the non-abelian 2-group structure between $\SU(N)_L \times \SU(N)_R$ 0-form symmetry and $\U_\text{{wind}}$ 1-form symmetry, with Postnikov classes $\hat{\kappa}_L=n$ and $\hat{\kappa}_R = -n$ \cite{Cordova:2018cvg}.

To summarise so far, we have shown that the IR mixed WZW term implies a non-trivial 2-group structure, with $\U$ as the 1-form part. In the following Sections, we will use this 2-group structure as a guide to build a UV completion for this EFT.

\section{From infrared to ultraviolet: a no-go theorem} \label{sec:no-go}

One might try to UV complete this `product coset' model with a QCD-like strongly coupled gauge theory, with two types of quark field that confine at different scales to give the two types of pions in the IR.
We have already seen that such a UV completion does not fix the mixed WZW term coefficient by anomaly matching, because there is no non-abelian mixed anomaly in the UV. Now we will show something even stronger, which is that the mixed WZW term in the IR is in fact {\em inconsistent} with the QCD-like dark sector completion. We will then build a UV completion that does work in \S \ref{sec:proper-completion}.

To see why a QCD-like dark sector cannot generate this mixed WZW term, let's have a concrete model in mind.
To get the symmetry breaking pattern $\SU(2)_D \to \U_D$, a candidate dark dynamics is 
$\SO(N_c)$ gauge theory with two flavours of fundamental dark quark. Including also the QCD part, which takes the form of an $\SU(n_c)$ gauge theory acting on fundamental quarks, and allowing for other matter fields transforming in linear representations under both $\SO(N_c)$ and $\SU(n_c)$ that communicate weakly between the two sectors,
there is no continuous $\U^{[1]}$ 1-form symmetry in this phase (although there is a $\Z_2^{[1]}$ 1-form symmetry associated with the gauged $\SO$ group).

But the absence of a 1-form symmetry means the current algebra for the 0-form QCD symmetries does not close; recall the Postnikov class appearing in the 2-group current algebra relations (\ref{eq:2grp-algebra-L}, \ref{eq:2grp-algebra-R}) is integer-quantized, and so preserved under RG flow.
Put in a more general context, the problem is that the IR theory (with non-zero mixed WZW term) encodes a non-trivial {\em extension} between the flavour symmetry and the 1-form symmetry, manifest in the fibration
\begin{equation}
    B\U \hookrightarrow \mathbb{G}_{\text{IR}} \to \SU(N)_L \times \SU(N)_R\, 
\end{equation}
being topologically non-trivial (where here $\mathbb{G}_{\text{IR}}$ denotes the infrared 2-group symmetry we have detected). This means that, if only the 0-form $\SU(N)^2$ flavour symmetry is there in the UV, then the 1-form symmetry cannot be emergent and end up twisted in such a non-trivial fibration.\footnote{The same argument `against infrared emergence' should apply in other contexts, not just for 2-group symmetry but for any infrared symmetry that is a non-trivial extension. We thank Y. Tachikawa for raising this point.}
As a result, this kind of dynamics {\em cannot} possibly UV complete our IR EFT with mixed WZW term, unless the UV also breaks the 0-form flavour symmetry. 

We thus establish a `no-go theorem':
\begin{quote}
        In the absence of any explicit symmetry breaking, the sigma model on $G/H \times \SU(2)/\U$ with non-zero mixed WZW term cannot be UV completed by a non-abelian gauge theory with semisimple gauge group.
\end{quote}
A valid UV completion that preserves the 0-form QCD flavour symmetries {\em must}, at the very least, possess a non-trivial 1-form symmetry that can close the 2-group operator algebra. 

One can refine this picture to consider UV theories in which both the 0-form flavour symmetry and the winding number 1-form symmetry are emergent in the infrared. The {\em 2-group emergence theorem}\footnote{This 2-group emergence theorem has been applied, for instance, to study models for unification in~\cite{Cordova:2022qtz} and, perhaps more closely to the present work, to theories with axions in~\cite{Brennan:2020ehu,Choi:2022fgx}.} 
of Ref.~\cite{Cordova:2018cvg} then implies the scale hierarchy $\Lambda_{\text{flavour}} \lesssim \Lambda_{\text{1-form}}$, where $\Lambda_{\text{flavour}}$ ($\Lambda_{\text{1-form}}$) denotes the scale where the 0-form QCD flavour symmetry (1-form symmetry) emerges. Assuming the 0-form symmetry remains an exact symmetry of the UV, as in a QCD-like completion with explicit symmetry breaking sources turned off, corresponds to the limit $\Lambda_{\text{flavour}} \to \infty$. 

One might object, however, that the Landau pole associated with the abelian gauge field means that the weakly coupled theory we  construct in \S \ref{sec:proper-completion} is not really UV `complete'. In \S \ref{sec:Xray} we sketch how that theory could be further embedded in an asymptotically-free gauge theory, and see how this explicitly entails breaking of the flavour symmetry at finite $\Lambda_{\text{flavour}}$.

\section{From infrared to ultraviolet: a weakly-coupled completion} \label{sec:proper-completion}

Informed by this observation, we propose that the mixed WZW term
arises from a particular coupling between QCD and scalar electrodynamics that 
enables a non-trivial 2-group structure. The 0-form part of the 2-group structure is the
$\SU(N)_L$ (and, separately, $\SU(N)_R$) flavour symmetry, while
the 1-form part is the $\U$ magnetic symmetry associated with the $\U$ gauge symmetry 
of the scalar electrodynamics. 

We stress that it is crucial that the extra gauge symmetry be abelian to bestow us with a candidate 1-form symmetry, with which we can try to close the 2-group current algebra in the UV and thence match the IR symmetries encoded in the mixed WZW term.

\subsection{The UV phase: QCD coupled to scalar electrodynamics}
\label{sec:QCD+SED}

The particular scalar electrodynamics (SED) that we will couple to the QCD sector consists 
of two complex scalar fields $\phi_1$ and $\phi_2$, both coupled to the $\U$ gauge field 
$b$ with charge $+1$. The dynamics of our SED is governed by the Lagrangian
\begin{equation}
 \mathcal{L}_{\text{SED}} =  -\frac{1}{4e^2} (db)^2 + \sum_{i=1}^2 |(\partial-i b) \phi_i|^2 + m^2 \sum_{i=1}^2 |\phi_i|^2 + \lambda \sum_{i=1}^2 |\phi_i|^4,
\end{equation}
where $e$ is the gauge coupling, $m^2$ is the mass squared parameter for the scalars (of as-yet unfixed sign), and $\lambda$ is the parameter for the quartic potential. 
Without the gauging, the scalar potential would have an $\mathrm{O}(4)$ accidental symmetry, exactly like the Higgs sector of the electroweak theory with gauge fields turned off. When the $\U$ is gauged, akin to gauging hypercharge in the electroweak theory, the remaining accidental global symmetry is reduced to an $\SU(2) \subset \mathrm{O}(4)$.\footnote{ In the case of the electroweak theory, but not here, this $\SU(2)$ is of course also gauged.}

We then couple the quarks to the SED sector through the $\U$ gauge field, which we also take to have charge $+1$ for now.\footnote{We shall consider the generalisation in which the quarks and scalars have different charge, which actually introduces topological complications, in \S \ref{sec:scalar_variation}. }
The full Lagrangian is then given by
\begin{align}
  \mathcal{L}_{\text{full}} &= \mathcal{L}_{\text{QCD}} + \mathcal{L}_{\text{SED}} + \mathcal{L}_{\text{int}},\label{eq:full-uv-lagrangian}\\
  \intertext{where}
  \mathcal{L}_{{\text{QCD}}} &=  -\frac{1}{2g^2} \Tr f_{\mu\nu}f^{\mu\nu} + \sum_{i=1}^{N} i \overline{\Psi}_i \left( \slashed{\partial} -i \slashed{a} \right) \Psi_i,\\
  \mathcal{L}_{\text{int}} &= \sum_{i=1}^{N} \overline{\Psi}_i\gamma^{\mu}b_\mu\Psi_i
\end{align}
Here, $a$ denotes the $\SU(n_c)$ colour gauge field, and $f= d a -i a\wedge a$ its field strength.

\subsection{Symmetries and anomalies of the UV theory}

The faithfully acting global symmetry of this theory appears to be
\begin{equation}
\label{eq:globsym}
    G_{\text{glob}} = \frac{\U_q \times \SU(N)_L \times \SU(N)_R\times \SU(2)_\phi}{\Z_{n_c} \times \Z_N \times \Z_2} \times \U^{[1]}_m\;,
\end{equation}
To see this, let us first enumerate 0-form symmetries. As in the usual massless QCD, there are flavour 0-form symmetries $\SU(N)_L$ and $\SU(N)_R$ that acts on the left-handed and right-handed components of $\Psi$ independently. As already remarked, there is an $\SU(2)$ flavour symmetry for the scalars acting by
\begin{equation}
    \SU(2)_\phi: \qquad \phi_i \mapsto U_i^{\phantom{i}j} \phi_j, \quad U \in \SU(2)\,.
\end{equation}
Lastly, despite coupling the quarks to a $\U$ gauge symmetry, the $\U_q$ quark symmetry
\begin{equation}
      \U_q : \quad \Psi_i \mapsto e^{i \alpha} \Psi_i\,,
\end{equation}
remains a good global symmetry. This is because the $\U$ gauge symmetry acts on both the quarks and the scalars, leaving the rotations on the quarks alone independent. This matches our accounting for the gauged $\U$ in the scalar part above.

The quotient by $\Z_{n_c}$ in Eq. \eqref{eq:globsym} reflects the fact that the subgroup $\Z_{n_c} \subset \U_q$ can be rotated away by a $\SU(n_c)$ gauge transformation. On the other hand, the $\Z_N$ quotient is there to avoid double counting, because the action of the $\Z_N$ subgroup of the diagonal $\SU(N) \subset \SU(N)_L \times \SU(N)_R$ is the same as the action of the $\Z_N$ subgroup of $\U_q$. The quotient by $\Z_2$ is due to  the fact that the centre of $\U_q \times \SU(2)_\phi$ coincides with a $\Z_2$ subgroup of the $\U$ gauge group, and thus can be gauged away.

In addition to these 0-form symmetries, there is a magnetic 1-form symmetry $\U_m^{[1]}$ acting on 't Hooft line defects, whose conserved charges measure the magnetic fluxes of these defects. The symmetry inevitably arises in a $\U$ gauge theory. Because the $\U$ field strength $h:=db$ is closed, $dh=0$, it naturally defines a 2-form current $j_m^{(2)} := \star h/2\pi$ which is conserved,
\begin{equation}
    d \star j_m^{(2)} = \frac{dh}{2\pi} = 0\,,
\end{equation}
even before imposing the equations of motion. It is therefore an example of a topologically-conserved 1-form symmetry.
Notice that this enjoys the same status as the topologically-conserved 2-form $j_{\text{wind}}^{(2)}$ that we previously identified in the IR sigma model.

This is not the whole story. The magnetic 1-form symmetry and the flavour symmetries form a non-trivial 2-group structure. The Postnikov class which characterises the structure can
be read-off directly from the associated anomaly polynomial of the
theory once we turn on the background gauge fields $A_{L/R}$ for
$\SU(N)_{L/R}$, with corresponding field strength 2-forms $F_{L/R}$. The degree-6 anomaly polynomial is given by 
\begin{equation} \label{eq:Phi6}
\Phi_6 = \frac{n_c}{3!} \frac{1}{(2\pi)^3} \left[ \Tr F_L^3 - \Tr F_R^3 \right] + \frac{n_c}{2} \frac{h}{2\pi} \left[ \Tr \left( \frac{F_L}{2\pi} \right)^2 - \Tr \left( \frac{F_R}{2\pi} \right)^2 \right]
\end{equation}
The first term represents the usual 't Hooft anomaly for the
$\SU(N)_{L/R}$ chiral global symmetries, that is matched in the IR by the familiar WZW term of pure QCD constructed from $\Tr \left( g^{-1} d g \right)^5$. 

The second term represents
the `operator-valued mixed anomalies' between $\SU(N)_{L/R}$ and the
$\U$ gauge symmetry, and can be properly interpreted as intertwining the
$0$-form symmetries $\SU(N)_{L/R}$ with the 1-form magnetic symmetry to form a non-trivial 2-group structure \cite{Cordova:2018cvg}. The Postnikov classes characterising these
2-group structures are given by the anomaly coefficient as
\begin{equation}
\hat{\kappa}_{L} = -\hat{\kappa}_{R} = n_c\,.
\end{equation}
Comparing with the 2-group structure enshrined by the mixed WZW term in the IR, which recall was captured by Postnikov classes in Eq. (\ref{eq:Post_IR}), suggests that matching of the global symmetries fixes the coefficient of the mixed WZW term in the IR:
\begin{equation}
    n=n_c\, ,
\end{equation}
in very close analogy to the coefficient of the ordinary WZW term that is fixed by the 't Hooft anomaly! 

Of course, we have not yet shown that this theory actually UV completes the IR sigma model with mixed WZW term; all we have so far shown is that the generalised global symmetries we identified (and the 't Hooft anomalies, as inherited from the pure QCD part) all match. In the next few Sections, we explicitly follow the RG flow, starting from this UV theory, to show how we arrive at the mixed WZW terms after going through two phase transitions.

\subsection{The Higgs phase} \label{sec:Higgs-phase}

To begin, we first define the parameters in the scalar sector (with a choice of sign for the quadratic term) so
that the potential can be written
\begin{equation}
V(\phi_i) = \lambda \left( |\phi_i|^2 - v^2 \right)^2,
\end{equation}
so that the scalars acquire a non-zero vacuum expectation value (VEV). Without turning on the $\U$ gauge field $b$, the vacuum manifold would be $S^3$,
given by the minima of the potential. This reflects the symmetry
breaking pattern $\mathrm{O}(4) \to \mathrm{O}(3)$. Taking the $\U$ quotient from the gauge group reduces 
the vacuum manifold down to $S^2$.

To simplify matters it is helpful to stagger the phase transitions in the QCD sector and the scalar sector, so that there are two distinct matching steps to trace out. 
To that effect, we assume a large separation of scales, 
\begin{equation}
    \Lambda_{\mathrm{QCD}} \ll v\, ,
\end{equation}
so that the Higgsing occurs first. The idea is that in this first Higgsing step we integrate out the heavy degrees of freedom (a heavy gauge field and a radial scalar mode) to get an intermediate effective description of the remaining light scalars coupled to quarks. Then we follow the RG flow through the subsequent chiral symmetry breaking transition by which quarks and gluons give way to pion degrees of freedom. We assume that, triggered by the flow to strong coupling at the low scale, chiral symmetry breaking and confinement occurs in the QCD sector more-or-less unaffected by the weak coupling to the dark sector that is mediated by the abelian gauge field.

\subsubsection{Derivation of the mixed WZW term take 1: local form }

To get a feel for how the mixed WZW term arises, we shall first work locally to derive the local approximation to the mixed WZW term given in Eq. \eqref{eq:pert-mixed-WZW}, using the gauge fixing procedure familiar from the electroweak theory.\footnote{As anticipated above, the scalar QED part of our Lagrangian corresponds precisely to the electroweak theory describing the complex Higgs doublet, but with the $\SU(2)_L$ gauge coupling turned off.}
In unitary gauge, and taking the unbroken $\U_\phi$ subgroup to be generated by $\sigma^3/2\in \mathfrak{su}(2)_\phi$, 
we expand the doublet of complex scalar fields $\phi(x) := \left(\phi_1(x), \phi_2(x)\right)^T$ 
around a minimum of the potential as
\begin{equation}
\label{eq:unitary-gauge-expansion}
    \phi(x) =  e^{\frac{i}{2 f_D} \chi^i(x) \sigma^i} \begin{pmatrix}
0 \\
v + \frac{\rho(x)}{\sqrt{2}}
\end{pmatrix}\, ,
\end{equation}
where here the index $i$ runs only from 1 to 2, {\em i.e.} over the broken generators. The $\chi^3$ Goldstone mode has, in this gauge, been eaten to become the longitudinal mode of the Higgsed abelian gauge field $b$. For simplicity, we further assume a limit $m^2 \gg v^2$ (by taking $\lambda \gg 1$) to decouple the radial mode $\rho$, which we henceforth neglect, to obtain a non-linear sigma model description of the scalar sector in this Higgsed phase.

We now consider the effective field theory valid at energies $E$ in the intermediate r\'egime,
\begin{equation}
    \Lambda_{\mathrm{QCD}} \ll E \ll v\, .
\end{equation}
This means we can integrate out the heavy gauge field $b$, whose mass is order $v$ (assuming an order-1 gauge coupling). We do so at tree-level by setting $b$ to its classical equations of motion, and we work to leading order in the derivative expansion $\partial_\mu / v \sim E/v$, which is here equivalent to neglecting the kinetic term for $b$ in the equation of motion.

Then, the relevant terms in the Lagrangian that
involve $b$ come from the kinetic terms for $\phi_i$ and $\Psi_i$, which read
\begin{equation}
\mathcal{L}_{\mathrm{UV}} \supset\,\, b_{\mu}j_q^{\mu} + |(\partial- ib)\phi_i|^2 \supset v^2 b^2 + b_{\mu} j_{\phi}^{\mu} + b_{\mu} j_q^{\mu}, 
\end{equation}
where the quark and scalar currents are given locally by the 1-forms
\begin{align}
    j_q &= \overline{\Psi}_i \gamma_{\mu} \Psi_i \,dx^{\mu}\,,\\
    j_\phi &= \frac{v^2}{2f_D^2} \epsilon_{ij} \chi^i d \chi^j\,, \qquad i,j=1,2\, .
\end{align}
We emphasize that the latter equation is derived under the assumption that we're in a coordinate patch in the vicinity of the origin $\chi^1=\chi^2=0$, and does not necessarily hold away from such a patch.
The leading order equations of motion then give 
\begin{equation}
b = -\frac{1}{2v^2}(j_{\phi} + j_q) + \dots \, ,    
\end{equation}
so after integrating out $b$ we generate effective mass dimension-6 operators of the form
\begin{equation}
\mathcal{L}_{\mathrm{EFT}} \supset -\frac{1}{4v^2} \left( j_{\phi} + j_q \right)_\mu \left(  j_{\phi} + j_q  \right)^\mu\, ,
\label{eq:L-EFT}
\end{equation}
which is the leading order (and local) result of our intermediate EFT matching step.
The cross-term will eventually match onto our mixed WZW term. 

Now consider flowing further into the deep IR, {\em i.e.} to energy scales
\begin{equation}
    E \ll \Lambda_{\text{QCD}}\, .
\end{equation}
The QCD chiral symmetry $\SU(N)_L \times \SU(N)_R$ breaks spontaneously down to its diagonal
subgroup $\SU(N)_V$ due to the non-vanishing chiral condensate. The QCD part of the resulting sigma model description is known as
the chiral Lagrangian, for which the leading order action is
\begin{equation}
  S_{\chi}[g] = \int_{\Sigma_4} \frac{f_{\pi}^2}{4} \Tr \left( \partial_{\mu}g^{\dag} \partial^{\mu}g \right) + n_c \Gamma[g]\,,
\end{equation}
where the dynamical field
\begin{equation}
  g(x) = \exp \left( \frac{2i}{f_{\pi}} \pi_a (x) t_a\right) \in \SU(N) \cong \frac{\SU(N)_L \times \SU(N)_R}{\SU(N)_V}
\end{equation}
describes the pions $\pi(x)$, with the pion decay constant $f_{\pi}$ and where $t^a$ are the generators of $\SU(N)$. We include the
usual WZW term $i n_c \Gamma[g]$, with
\begin{equation}
\Gamma[g] = \frac{1}{240 \pi^2} \int_{\Sigma_5} \Tr \left( g^{-1} d g \right)^5\,, \quad \partial \Sigma_5 = \Sigma_4\,,
\end{equation}
which is needed to match the 't Hooft anomalies in the $\SU(N)^2$ QCD flavour symmetry.

Most important for us, however, is what becomes of the interaction term between
the quark and the scalar sectors, which is the cross-term in the effective coupling in \eqref{eq:L-EFT}, namely
\begin{equation}
  \mathcal{L}_{\text{int}} = -\frac{1}{2v^2} j_{q,\,\mu}  j_{\phi}^\mu\,.
\end{equation}
Due to confinement, the quarks now combine into
baryons which carry $\U_q$ charge. Because one baryon consists of $n_c$ quarks, the baryon
current $j_B$ is given in terms of the quark current $j_q$ by
\begin{equation}
j_q = n_c j_B\,,
\end{equation}
with, importantly, a relative factor of $n_c$ appearing in the normalisation of these currents.
In the chiral Lagrangian, the baryons can be identified as solitons (\`a la Skyrme~\cite{Skyrme:1961vq}) formed from the
pion fields. In this description, the baryon number current $j_B$ is then given in terms of $g$ by the topologically conserved form~\cite{Balachandran:1982dw,Witten:1983tx}
\begin{equation}
\star j_B = \frac{1}{24\pi^2} \Tr \left( g^{-1}dg \right)^3 = \frac{1}{24\pi^2}\frac{2}{f_\pi^3}f_{abc}d\pi_a \wedge d\pi_b \wedge d\pi_c + \mathcal{O}(\pi^4)\, ,
\end{equation}
and the integral of $\star j_B$ measures the baryon number of a pion field configuration.

The cross interaction term in our EFT Lagrangian becomes
\begin{equation}
    \mathcal{L}_{\text{int}} = -\frac{n_c}{2v^2} j_{B,\,\mu} j_{\phi}^\mu\,.
\end{equation}
Expanding both currents in terms of the pion fields $\pi_a(x)$ and the sigma model fields $\chi_i(x)$, and integrating by parts to move a derivative,
we obtain the local Lagrangian
\begin{align}
    \mathcal{L}_{\text{int}} 
    = \frac{n_c\epsilon^{\mu\nu\rho\sigma}}{48\pi^2 f_D^2 f_\pi^3} f_{abc} \epsilon_{ij} \pi_a \partial_\mu \pi_b \partial_\nu \pi_c \partial_\rho \chi_i \partial_\sigma \chi_j + \mathcal{O}(\pi^4 \chi^2, \pi^3 \chi^3)\, .
\end{align}
This matches the local form of the mixed WZW term given in Eq. \eqref{eq:pert-mixed-WZW}, with the IR coefficient fixed to be the number of colours in the QCD sector, $n=n_c$;
precisely as the 2-group symmetry matching argument at the end of \S \ref{sec:QCD+SED} suggested.

However, the na\"ive leading order EFT matching we have just demonstrated is not quite sufficient in this scenario, precisely because of the fundamentally topological nature of this interaction.
By working with only the local form of the currents and Lagrangian terms (enforced by our use of local coordinates $\{\chi^i\}$), which we did to elucidate
the perturbative physics as clearly as possible, we have ignored important and non-trivial topological data concerning the vacuum manifold $S^2$. In particular, one could not with this formalism hope to show that the global form of the topological term must be written in terms of $\Vol$, and that the term therefore requires an extension to an auxiliary 5d bulk.
In the following Subsection we will patch up our derivation, to give a globally-valid account of the EFT matching onto this mixed WZW term.

\subsubsection{Derivation of the mixed WZW term take 2: global form}

To make the topological information about the vacuum manifold (in particular, its non-trivial second homology and second homotopy groups, and associated winding number) manifest in the final 
mixed WZW, we have to derive it with a less direct implementation of gauge fixing, that does not require a local expansion of the underlying scalars $\phi_i$ in terms of the dark pion fields $\chi^i$.

First, instead of the unitary gauge 
used in Eq. \eqref{eq:unitary-gauge-expansion}, let us now expand the scalar fields around the vacuum manifold as
\begin{equation}
\phi_i = z_i + h_i, \qquad |z_1|^2 + |z_2|^2 = v^2,
\end{equation}
where $z_1, z_2$ (subject to the constraint) describe the vacuum manifold $S^3$ of radius $v$, and
$h_i$ are the transverse fluctuations that give rise to the single radial mode $\rho(x)$ after gauge fixing. 
As in the usual spontaneous symmetry
breaking story, the potential $V(\phi)$ then tells us that the radial
modes are massive and can be integrated out, while the vacuum
manifold's degrees of freedom are massless, corresponding to the
NGBs.

The effect of gauging the $\U$ subgroup is that the affine coordinates $(z_1, z_2)$ become a pair
of homogeneous coordinates $[z_1:z_2]$, where we identify\footnote{We must also identify 
$\Psi \sim e^{i\alpha} \Psi$ at the same time since the quarks are also charged under
this $\U$ gauge symmetry.} 
\begin{equation}
    (z_1, z_2) \sim e^{i\alpha} (z_1, z_2)
\end{equation}
These homogeneous coordinates
form a familiar description of the 1-dimensional complex projective
space $\mathbb{C}P^1$, which is topologically a 2-sphere $S^2$.  
We have now transitioned from a geometric description of spontaneous symmetry breaking (of a global symmetry) to a geometric description of the Higgs mechanism,
where one would-be NGB morphs into the longitudinal mode of the gauge field and gets integrated out.
The scalar electrodynamics sector is then effectively described by 
a sigma model whose target space is properly identified (globally) with the manifold  $\mathbb{C}P^1$, which we will informally call 
the $\mathbb{C}P^1$-model.

We can then repeat the steps in the previous Subsection and integrate out massive fields to
obtain the EFT 
\begin{equation}
\mathcal{L}_{\mathrm{EFT}} \supset -\frac{1}{4v^2} \left( j_{\phi} + j_q \right) \wedge \star \left(  j_{\phi} + j_q  \right) 
\end{equation}
at an intermediate energy scale between $v$ and $\Lambda_{\text{QCD}}$. 
The only difference here is the form $j_\phi$ takes.
In the current gauge fixing scheme, we appear to have 
\begin{equation} \label{eq:jphi_z}
j_{\phi} = -i \left( dz_i^{*} z_i - z_i^{*} dz_i \right)\,.
\end{equation}
Again, flowing further down the RG flow replaces the QCD quark-gluon description with the chiral Lagrangian, 
with the quark current $j_q$ replaced by $n_c j_B$.
Just like in the previous Subsection, one might be tempted to write the cross interaction term as
\begin{equation}
  S_{\text{int}} \stackrel{?}{=} -\frac{n_c}{2v^2} \int_{\Sigma_4} \star j_B \wedge j_{\phi}\,,
    \label{eq:IR-proposed}
\end{equation}
but that would be wrong!

The problem with the proposed interaction \eqref{eq:IR-proposed}
lies in the fact that it is not gauge invariant. This is because on $S^2$, the object 
\begin{equation}
    \mathcal{A}_\phi:=\frac{1}{2v^2}j_{\phi}
\end{equation}
is not a globally-defined form, but rather behaves like a gauge connection. This becomes evident in our new description via homogeneous coordinates $\{z_i\}$, which still suffer from a gauge redundancy. To wit, we observe that under the $\U$ gauge transformation 
\begin{equation}
    (z_1,z_2) \to e^{i \alpha} (z_1,z_2), \qquad \Psi\to e^{i\alpha} \Psi
\end{equation}
that defines the homogeneous coordinates $[z_1:z_2]$ on $S^2$, the connection $\mathcal{A}_\phi$ obtained from \eqref{eq:jphi_z} transforms as
\begin{equation}
\label{eq:A-gauge-tranf-orig}
    \mathcal{A}_{\phi} \to \mathcal{A}_{\phi} - d \alpha\,.
\end{equation}
To emphasize the difference with our previous (necessarily local) description, the new description via homogeneous coordinates $\{z_i\}$ naturally covers the whole target space, but at the expense of having not fully fixed the gauge yet. And indeed one finds that the current $j_\phi$ is not gauge-invariant, but transforms (after an appropriate rescaling) as a connection. In contrast, things were fully gauged-fixed in the $\chi^i$-based formulae, but were necessarily restricted to a local patch near the origin, so not well-suited to studying field configurations that wind the $S^2$. 

Fortunately, there is a known path to proceed in such a situation, because the putative Lagrangian \eqref{eq:IR-proposed} behaves
exactly like a Chern--Simons term, which is similarly not gauge-invariant but which (iff properly-quantized) we know how to define rigorously
in terms of the field strength by going to one dimension higher.\footnote{
The exponentiated action here can also be defined as an invariant differential character~\cite{alexander1985differential,bar2014differential,Davighi:2020vcm} on the pion target space, the curvature of which is the globally-defined closed 5-form that we integrate in Eq. \eqref{eq:matched!}. This curvature form is analogous to the anomaly polynomial $\Phi_{d+2}$ in defining the Chern--Simons action in $d+1$ dimensions. From this perspective, the action requires quantized coefficient precisely because it cannot be expressed via a locally-defined 4-form, and can be seen without passing to an extra dimension by instead patching together locally defined forms using the tools of \v Cech cohomology~\cite{Wu:1976qk,Alvarez:1984es,Davighi:2018inx}, and demanding consistency.} 
The correct interaction term takes the form 
\begin{align}
  S_{\text{int}} &= -n_c \int_{\Sigma_5} d \left( \star j_B \wedge \mathcal{A}_{\phi} \right)\\
                 &= -n_c \int_{\Sigma_5} \star j_B \wedge d\mathcal{A}_{\phi} 
\end{align}
where we have used the fact that $j_B$ is a topological current, that is $d\star j_B =0$ off-shell.

All that remains is to show that $d \mathcal{A}_{\phi}$, which really denotes the curvature of the 1-form connection $\mathcal{A}_\phi$, is proportional to the volume form on the vacuum
manifold $S^2$. Recall that this vacuum manifold is described by the
pair of homogeneous coordinates $[z_1: z_2]$ satisfying
$|z_1|^2 + |z_2|^2 = v^2$. It can be shown (see, for instance, the book of Bott and Tu~\cite[\S 17]{Bott-Tu:1982}) that the volume
form $\Vol$ on this manifold, normalised so that
$\int_{S^2} \Vol =1$, can be written in terms of $z_0,z_1$ as 
\begin{equation}
\Vol = -\frac{i}{2\pi v^2} dz_i^{*} \wedge dz_i\,.
\end{equation}
On the other hand, we also find from Eq. \eqref{eq:jphi_z} that $dj_{\phi} = 2i dz_i^{*} \wedge dz_i$. So, as
promised, we obtain
\begin{equation}
d \mathcal{A}_{\phi} = -2\pi \Vol\,.
\end{equation}
Note that this is consistent with $\int_{S^2} \Vol = 1$ although 
$\Vol$ appears to be an exact form, because $\mathcal{A}_\phi$ is really a connection 1-form (not a globally-defined 1-form) and $d\mathcal{A}_\phi$ is shorthand for its curvature. Note also that the first Chern number associated to this connection is correctly quantized, {\em viz.} $c_1=\int_{X_2} \frac{ d\mathcal{A}_\phi}{2\pi} \in \Z$ for $X_2$ any 2-cycle in $\Sigma_4$, and where we use the same notation $d\mathcal{A}_\phi$ for the pullback of $d\mathcal{A}_\phi$ under $z_i: \Sigma_4 \to S^2$.

After we replace $d\mathcal{A}_\phi$ in terms of $\Vol$, the cross interaction term now reads
\begin{align}
  S_{\text{int}} &=\, 2\pi n_c \int_{\Sigma_5} \star j_B \wedge \Vol \\
  &=\, 2\pi n_c \int_{\Sigma_5} \frac{1}{24\pi^2} \Tr \left( g^{-1}dg \right)^3 \wedge \Vol\,, \label{eq:matched!}
\end{align}
reproducing precisely (and globally) the mixed WZW term that we are after, with all the topological data manifest. 
Again, we see that the WZW coefficient is given by
\begin{equation}
    n = n_c \nonumber
\end{equation}
More generally, if in the UV the abelian gauge field couples to the quarks with charge $X_q$ (but still charge $+1$ to the scalars), then the mixed WZW coefficient $n$ becomes $X_q n_c$. In this case, the Postnikov class appearing in the 2-group structure is also modified to $X_q n_c$ (as can be seen by straightforwardly adapting the argument of \S \ref{sec:QCD+SED}). The case of non-minimal scalar charge $X_\phi$ is slightly different, as we discuss in \S \ref{sec:scalar_variation}. 

\subsubsection*{A tree-level exact result}

We pause to make some further comments before continuing. First, we emphasize that it is extremely non-generic that we were able to match an interaction involving QCD through the chiral symmetry breaking transition into the chiral Lagrangian! This was only possible because the interaction with QCD was via baryon number current, which is robustly identified with a topologically conserved current in the IR. Likewise on the dark side, the coupling of the abelian gauge field is special, in that the connection 1-form associated to this coupling is a topologically non-trivial connection. When combined, these two special features contrive to mean that one obtains a {\em bona fide} quantized topological term --- from integrating out a weakly coupled abelian gauge field at tree-level. 

Due to the integer quantization of this coefficient (in appropriate units, which absorb the factors of $f_\pi$ and $v$ if we are using the local coordinate expressions), it follows that this tree-level matching result for the coefficient ought to be exact. The operator can only have an integer coefficient for consistency (as can be inferred purely from the low-energy EFT), and any corrections to this leading term, in the form of a perturbative series in the $\R$-valued couplings, could not maintain this integrality as the couplings run under RG flow. This is exactly analogous to the non-renormalisation of the chiral anomaly. The difference is that for the anomaly the result is 1-loop exact, whereas here the leading order term is already there at tree-level. 

This situation is reminiscent of how anomalies match for the Schwinger model ({\em i.e.} the 2d theory of a Dirac fermion coupled to a $\U$ gauge field) not under RG flow but across the bosonization duality. In that case, the mixed anomaly between the vector and the axial $\U$ symmetries on the fermionic side, which arise at 1-loop from the chiral fermion path integral, is matched by the mixed anomaly between the shift and winding symmetries on the bosonic side, which follows just from the tree-level equations of motion (see {\em e.g.} the lecture notes~\cite[\S7.5.6]{tong2018gauge}).

Another way to see this is that the coefficient $n$ appears in the 2-group current algebra where it is necessarily integer-quantized. It is this quantization that justifies the `symmetry matching' across RG flows (akin to the more familiar `anomaly matching') that we used in the beginning to suggest the no-go theorem of \S \ref{sec:no-go}.

\subsection{Phase structure of QCD coupled to scalar electrodynamics}

In this Section, we briefly describe how the IR dynamics of QCD coupled to scalar electrodynamics, namely the UV theory described in §\ref{sec:QCD+SED}, changes as we vary the dimensionless parameter
\begin{equation}
    \mu^2:= \frac{m^2}{\Lambda_{\mathrm{QCD}}^2}
\end{equation}
while keeping the scalar quartic coupling $\lambda \sim \mathcal{O}(1)$ fixed,
where recall $m^2$ is the mass-squared parameter for the scalars $\phi_i$ and $\Lambda_{\mathrm{QCD}}$ is the strong coupling scale for the QCD sector at which the chiral symmetry breaking occurs.
\begin{itemize}
    \item[(a).] $\bm{ \mu^2 \to +\infty}$. In this phase, the scalars are extremely massive and can be integrated out entirely. The remaining theory is QCD with gauge group $U(n_c)$ coupled to $N$ fundamental quarks. The gauged $\U$ symmetry, which in the UV acted to rotate both scalars and quarks, in the IR acts only as a gauging of baryon number. There is no additional non-trivial $\U$ global symmetry remaining once the scalars are lifted.
    \item[(b).] $\bm{0<\mu^2<1}$. In this case, chiral symmetry breaking occurs in the QCD sector while the scalars remain dynamical, and the photon (from gauging the mediator $\U$) remains massless. The IR theory consists of QCD mesons and baryons, with the baryons coupled weakly via the photon to two complex scalars of mass $m_\phi \sim \sqrt{\mu^2 \Lambda_{\mathrm{QCD}}^2} < \Lambda_{\mathrm{QCD}}$. Going to the deep IR, one would also integrate out the baryons and scalars and obtain a theory of weakly-interacting massless mesons and photons.
    \item[(c).] $\bm{0>\mu^2>-1}$. Here the scalars condense, triggering the symmetry breaking pattern described in the main text, but the QCD chiral symmetry breaking transition occurs first. The deep IR phase is the same as that described in the main text, namely of QCD and dark pions coupled to each other via the mixed WZW term. But, if we assume a scale separation ($|\mu^2|\ll 1$), one can study the EFT describing the intermediate phase. The situation is now `reversed' to that studied in \S \ref{sec:Higgs-phase}, in that it features QCD pions and baryons, with the baryon current coupled via the massive (but still dynamical) abelian gauge field to the pair of dark complex scalars. In this intermediate phase, the 2-group structure is matched by a term
        \begin{equation}
            S_{\text{int}} = 2\pi n_c \int_{M_5} \frac{\Tr \left( g^{-1}dg \right)^3}{24\pi^2} \wedge \frac{h}{2\pi}\, ,
        \end{equation}
    where recall $h=db$ is the $\U$ field strength.
    This ordering of the phase transitions could just as well have been used to rigorously derive the emergence of the mixed WZW term in the deep IR.
    \item[(d).] $\bm{\mu^2 \to -\infty}$. This is the limit discussed in the main text, which follows the RG flow described in \S \ref{sec:Higgs-phase}.
\end{itemize}

\subsection{Deeper into the UV} \label{sec:Xray}

Because of the Landau pole associated to the abelian gauge field, the short-distance phase we have set out is arguably not a true UV completion. To address this, we here sketch how the abelian gauge theory presented could be further UV completed into a semi-simple gauge theory. But of course, in accordance with the no-go theorem of \S \ref{sec:no-go}, this necessitates the quark flavour symmetry be only emergent in the abelian gauge theory phase.

The idea will be that the $\SU(n_c) \times \U$ gauge symmetry emerges from spontaneously breaking a larger $\SU(n_c+1)$ gauge symmetry that is linearly realised in the deep UV. Let us embed $\SU(n_c)$ as the upper left $n_c \times n_c$ block in the defining representation of $\SU(n_c+1)$, and take $\U$ to be generated by the traceless matrix $\mathrm{diag}(1,\dots,1,-n_c)\subset \mathfrak{su}(n_c+1)$. 
The key consideration will be the fermion sector, its flavour symmetries, and mixed anomalies with $\U$. 
Recall that we have $N$ flavours of left- and right-handed quark fields $\Psi_i$ in the fundamental representation of $\SU(n_c)$ and with unit $\U$ charge, as well as a pair of complex scalars that are $\SU(n_c)$ singlets and with unit $\U$ charge also. Now extend this field content by $N$ flavours of left- and right-handed `lepton' fields, that are $\SU(n_c)$ singlets but carry $\U$ charge $-n_c$. The global symmetry of the fermion sector is now enhanced from the $\SU(N)_L^q \times \SU(N)_R^q$ quark symmetry of before (now with $q$ for `quark' labels in superscript), that participated in the 2-group structure matched in the deep IR by the mixed WZW term, to a larger
\begin{equation}
    \SU(N)_L^q \times \SU(N)_R^q \times \SU(N)_L^l \times \SU(N)_R^l
\end{equation}
flavour symmetry. If we turn on independent background gauge fields for all four of these $\SU(N)$ factors, the anomaly polynomial contains a piece
\begin{equation} 
\Phi_6 = \frac{n_c}{2} \frac{h}{2\pi} \left[ \Tr \left( \frac{F_L^q}{2\pi} \right)^2 - \Tr \left( \frac{F_R^q}{2\pi} \right)^2 - \Tr \left( \frac{F_L^l}{2\pi} \right)^2 + \Tr \left( \frac{F_R^l}{2\pi} \right)^2 \right]\, .
\end{equation}
Note that the overall coefficient of $n_c$ comes from the sum over colour components in the case of the quark piece, while it comes from the ratio of $\U$ charges for the lepton piece. We see that, for the diagonal subgroup of flavour transformations
\begin{equation}
    G_{\text{diag}} := \SU(N)_L^{q+l} \times \SU(N)_R^{q+l}\, ,
\end{equation}
the 2-group coefficients vanish. Accordingly, symmetry matching poses no obstruction to a semi-simple unification pattern (for which there is no 1-form symmetry) whereby quarks and leptons are unified. Sure enough, the $\U$ charge assignment is chosen such that quarks and leptons package into the fundamental representation of $\SU(n_c+1)$.

Starting from this deeper UV theory and running down, we have that the separate quark and lepton flavour symmetries {\em emerge} at the scale where $\SU(n_c+1)$ is Higgsed down to $\SU(n_c)\times \U$, alongside the magnetic 1-form symmetry associated to the $\U$, and that these 0-form and 1-form global symmetries are fused into 2-group symmetry.

Of course, to write an explicit model the scalars must also be embedded in representations of $\SU(n_c+1)$, which necessitates the inclusion of additional coloured scalars, and a mechanism must be put forth for decoupling the extra scalars and the leptons, but we are content to postpone such considerations here.

\section{Variations of the scalar sector}\label{sec:scalar_variation}

In this Section we discuss three variations in the scalar sector of the theory.

\subsection{Non-minimal scalar charge} 


In the main text (\S \ref{sec:Higgs-phase}) we discussed the straightforward modification that follows from varying the abelian quark charge $X_q$; in the UV, the term in the anomaly polynomial responsible for the 2-group structure (and thus the Postnikov class) is simply rescaled by $X_q$, and this tracks all the way through the RG matching to rescale the coefficient of the WZW term. 

But what if the scalar charge is taken to be non-minimal, $X_\phi \neq 1$ (in minimal units, {\em i.e.} such that $\text{gcd}(X_q, X_\phi)=1$)? This na\"ively presents a puzzle, because the 2-group structure is not modified, depending only on chiral fermion representations in the UV theory, but the coefficient of the WZW {\em does} appear to be rescaled, this time by a factor $1/X_\phi$ that arises due to the rescaling of both the $b\phi_i\phi_i^\dagger$ vertex and of the gauge boson mass.
This scenario requires a slightly more careful analysis, because the non-minimal scalar charge alters the symmetry breaking structure induced by the scalars condensing. 

Firstly, the non-minimal scalar charge means there is a discrete $\Z_{|X_\phi|}\subset \U$ gauge symmetry that remains unbroken by the scalar condensate, which acts non-trivially on the quark fields and can in principle be detected via its holonomy.

Concerning the dark pions,
the target space remains a 2-sphere, but with a rescaled volume form, satisfying 
\begin{equation} \label{eq:Xflux}
    \int_{S^2} \Vol = X_\phi\,.
\end{equation}
To see this, recall that $\Vol = d \mathcal{A}_\phi$ where $\mathcal{A}_\phi = -i(d z_i^* z_i - z_i^* d z_i)$ behaves like a gauge connection. However, under the $\U$ gauge transformation $z_i \mapsto e^{i X_\phi \alpha} z_i$, with $\alpha$ being $2\pi$-periodic, we have,
\begin{equation}
    \mathcal{A}_\phi \mapsto \mathcal{A}_\phi - X_\phi d\alpha,
\end{equation}
instead of \eqref{eq:A-gauge-tranf-orig}. Therefore, the properly normalised $\U$ connection is $\mathcal{A}_\phi / X_\phi$ instead of $\mathcal{A}_\phi$. Consequently, the minimally quantised volume form is $\Vol / X_\phi$.

The WZW term can be derived in exactly the same way as in \S \ref{sec:Higgs-phase} by integrating out the massive $\U$ gauge field, with a couple of changes already mentioned above. Firstly, since the mass squared of the gauge field is now equal to $X_\phi^2 v^2$, the cross interaction term between $j_B$ and $j_\phi$ that arises from integrating out the gauge field is now inversely proportional to $X_\phi^2$:
\begin{equation}
    S_{\text{int}} = -\frac{n_c}{2X_\phi^2 v^2} \int \star j_B \wedge j_\phi\,.
\end{equation}
Secondly, because $\phi$ now has charge $X_\phi$ under the $\U$ gauge group, the current $j_\phi$ is rescaled: 
\begin{equation}
    j_\phi = X_\phi j_\phi^{\text{(old)}} = 2X_\phi v^2 \mathcal{A}_\phi\,.
\end{equation}
Putting the two together, we arrive at 
\begin{equation}
    S_{\text{int}} = 2\pi n_c X_q \int_{\Sigma_5} \frac{1}{24\pi^2} \Tr \left(g^{-1}dg\right)^3 \wedge \frac{\Vol}{X_\phi}.
\end{equation}
Despite the modified coefficient, because of the compensating modified flux relation \eqref{eq:Xflux} the closed 5-form in the integrand is still integral, meaning this topological term is still of course well-defined. Moreover, the 2-group current algebra is unchanged {\em i.e.} is independent of $X_\phi$. To see this, we should take the Noether current for the 1-form symmetry to be
$j^{(2)}_{\mathrm{wind}} := \star_4 \mathrm{Vol}_{S^2}/X_\phi$, which is integral, and so the minimal coupling term \eqref{eq:Bcoup} is now $S_{\text{coup}}=i\int_{\Sigma_4} \frac{\Vol}{X_\phi}\wedge B$. If we repeat the analysis of \S \ref{sec:WZW}, we see the Postnikov class of the 2-group remains $\hat\kappa_L=-\hat\kappa_R=n_c X_q$.
This is consistent with matching the 2-group symmetry, resolving our little puzzle.

\subsection{From 2-sphere to 2-torus?} 

One might wonder if it is special that the two `dark' pions live on $S^2$, or whether a similar story plays out for a non-linear sigma model on $\SU(N)\times K$, where $K$ is some other homogeneous space with $H_{\mathrm{dR}}^2(K)\neq 0$ so that there is a cohomologically non-trivial 2-form that can play the role of $\Vol$ in constructing the mixed WZW. In particular, one might consider the case $K=(\U\times\U)/\{\cdot\}=T^2$, {\em i.e.} with a pair of `axion-like' dark pions living on a 2-torus. Let $(\phi_1,\phi_2)\in [0,2\pi)^2$ denote coordinates on $T^2$ in this Subsection.

Even though a na\"ive cohomology-based classification of topological terms~\cite{DHoker:1994rdl} would suggest there is a mixed WZW for this coset, 
there is in fact no such term if we insist the NLSM has exact $\U \times \U$ global symmetry acting by translations on $\phi_{1,2}$.\footnote{We have in mind that the pions arise as Goldstones on $G/H$ following spontaneous symmetry breaking, starting from a $G$-invariant Lagrangian, and seek to construct the most general EFT consistent with symmetry. 
Alternatively, one can dispense with the global symmetry and view the scalar field theory as arising from a general non-linear sigma model, in which case there is no obstruction to defining the mixed WZW on $\SU(N)\times T^2$.}
(In contrast, the WZW term on $\SU(N) \times (\SU(2)_\phi/\U_\phi)$ studied above {\em is} invariant under exact $\SU(N)_L\times \SU(N)_R\times \SU(2)_\phi$.) A putative WZW term, which one might define precisely to be a differential character with non-vanishing curvature form $\omega_{d+1}$, is $G$-invariant iff $\omega_{d+1}$ satisfies the so-called `Manton condition'~\cite{Davighi:2018inx,Davighi:2020vcm}, which requires the contraction of $\omega_{d+1}$ with each vector field generating the $G$-action (we assume $G$ is connected) be an exact form (not just closed).
The putative 5-form $\omega \sim \Tr (g^{-1}dg)^3 d\phi_1 d\phi_2$ on $\SU(N)\times T^2$ violates this condition, because
\begin{equation} \label{eq:manton}
    \{\iota_{\partial_{\phi_1}}\omega, \iota_{\partial_{\phi_2}}\omega \}\sim \{\Tr (g^{-1}dg)^3 d\phi_2, -\Tr (g^{-1}dg)^3 d\phi_1 \}
\end{equation}
are closed but not exact 4-forms. A classification of {\em invariant} topological actions using invariant (differential) cohomology~\cite{Davighi:2020vcm} tells us there is no such term in the IR consistent with the global $\U^2$ symmetry.\footnote{This failure of invariance is a higher-dimensional avatar of the fact that coupling a quantum particle on a torus to a homogeneous (classically) translationally-invariant magnetic field breaks translations down to a discrete subgroup, a fact noticed long ago by Manton~\cite{Manton:1985jm}. A similar phenomenon occurs~\cite{Davighi:2018xwn} for a non-minimal composite Higgs model proposed in~\cite{Gripaios:2016mmi}. }

We can also see the pathology from the point of view of 2-group symmetry.
While na\"ively there is still a locally conserved 2-form, associated to the closed volume form $\mathrm{Vol}_{T^2}=\frac{1}{4\pi^2} d\phi_1 \wedge d\phi_2$, there is in fact no topological charge associated to the putative 1-form symmetry because $\pi_2(S^1\times S^1)=0$, so there are no linking surfaces that are topologically 2-spheres through which to measure a monopole flux. So, there should be no line operators transforming non-trivially under this 1-form symmetry, with which to close the 2-group symmetry structure. 

Of course, if we do not restrict our EFT to building invariants  then there is nothing to prevent one from coupling QCD-like pions via the mixed WZW term to a pair of necessarily {\em pseudo} NGBs $\phi_{1,2}$ living on $T^2$.\footnote{With non-zero masses, a phenomenologist would refer to such pNGBs on $T^2$ as a pair of `axion-like particles', or ALPs.} The mixed WZW-like coupling just described provides a source of explicit $\U^2$ symmetry breaking, closely analogous to symmetry breaking by an ABJ anomaly, that contributes to the non-zero masses of the pNGBs. Under an axion shift symmetry $\phi_1 \to \phi_1 + \lambda_1$, the failure of the Manton condition encoded by \eqref{eq:manton} implies the mixed WZW action would shift by
\begin{equation}
S \mapsto S + \int_{\Sigma_4} \frac{\lambda_1}{2\pi} \frac{n}{24\pi^2}\Tr (g^{-1}dg)^3 \frac{d\phi_2}{2\pi}\,,    
\end{equation}
which mimics the non-invariance due to an ABJ anomaly but with the instanton density $F\wedge F$ replaced by the 4-form $\propto \Tr (g^{-1}dg)^3 d\phi_i$.

\subsection{From \texorpdfstring{$\mathbb{C}P^1$}{ℂP¹} to \texorpdfstring{$\mathbb{C}P^n$}{ℂPⁿ}}

Having discussed the torus and its subtleties, one might then ask if there are other cosets $K$, generalising $\SU(2)/\U$, for which the mechanisms we have described do go through. 

We remark that one set of examples is readily furnished by directly generalising the $\mathbb{C}P^1 \cong \SU(2)/\U \cong S^2$ model to 
\begin{equation}
    \mathbb{C}P^n \cong \frac{\SU(n+1)}{S[\U \times \mathrm{U}(n)]}    
\end{equation}
for $n \in \Z_{>1}$. The de Rham cohomology is 
\begin{equation}
    H^k_{\mathrm{dR}} (\mathbb{C}P^n) = \begin{cases}
        \R \qquad & k \text{~even~}, 0 \leq k \leq n\\
        0 \qquad & k \text{~odd~}\, ,
    \end{cases}
\end{equation}
and because $\mathbb{C}P^n$ is a symmetric space, the invariant forms are in 1-to-1 with cohomology classes. Moreover, because $G=\SU(n+1)$ is simple, the Manton condition discussed above reduces simply to requiring $G$-invariance of the differential form. Thus, a representative form $\Omega$ for the generator of the cohomology ring above, which can be identified with the K\"ahler form, can be used to construct a mixed WZW action of the form
\begin{equation}
    S[\Sigma_4] = \int_{X_5} \frac{m}{24\pi^2} \Tr (g^{-1}dg)^3 \wedge \Omega\, , \qquad n \in \Z\, , \,\, \partial X_5 = \Sigma_4\, .
\end{equation}
In homogeneous coordinates that generalise those introduced for $S^2 \cong \mathbb{C}P^1$ in \S \ref{sec:Higgs-phase}, defined as
\begin{equation}
    \Big\{ z_i \in \mathbb{C}^{n+1}\backslash \{0\} \quad | \quad \sum_i |z_i|^2=v^2, \quad z_i \sim e^{i\alpha}z_i \,\,\forall \alpha \in \R/2\pi \Z \Big\} \qquad  i=0,\dots n\, ,
\end{equation}
we can choose a representative 2-form to be given by the K\"ahler form (see {\em e.g.}~\cite{moroianu2007lectures})
\begin{equation}
    \Omega \sim  -\frac{i}{v^2} \sum_{i=0}^n dz_i \wedge dz_i^\ast \, ,
\end{equation}
appropriately normalised to have integer periods.
This term, for non-zero coefficient $m$, encodes non-trivial 2-group symmetry exactly as for the $n=1$ case studied at length in this paper.

The UV completion via QCD coupled to scalar electrodynamics should also generalise directly from $\mathbb{C}P^1$ to $\mathbb{C}P^n$, by passing from $2$ complex scalars $\phi_i$ to $n+1$ complex scalars, all coupled with the same charge to the abelian gauge field $b$ (that also couples universally to quarks in a vector-like fashion as before). This many-scalar model could provide an interesting variant of the dark matter portal mechanism proposed in~\cite{Davighi:2024zip}.

Yet further generalisation is possible if we allow $K$ to be a more general manifold $M$, not necessarily a coset arising from spontaneous symmetry breaking. The same mixed WZW term exists whenever $H^2_{\mathrm{dR}}(M)$ is non-trivial, with $\Omega$ similarly picked to be a 2-form representative of a non-trivial element in $H^2_{\mathrm{dR}}(M)$ with unit period. A partial UV-completion is provided by QCD coupled to a non-linear sigma model with target space being the line bundle $L\to M$ over $M$ whose Chern class is given by $[\Omega] \in H^2(M;\Z)$. We then gauge the diagonal between the $\U_q$ quark number symmetry and the $\U$ symmetry acting on the fibre of $L$.\footnote{We thank Y. Tachikawa for suggesting this more general approach.} The crucial non-trivial 2-group structure then arises from the mixed 't Hooft anomaly between $\U_q$ and $\SU(N)_{L/R}$ after this gauging \cite{Cordova:2018cvg} (the same phenomenon also happens when gauging finite groups with mixed 't Hooft anomalies \cite{Tachikawa:2017gyf}).

\section{Gauged version} \label{sec:gauging}

In this Section, we discuss the case in which an anomaly-free $\U$ subgroup of the QCD 0-form flavour symmetry $\SU(N)_L \times \SU(N)_R$ is gauged. This is a physically important scenario, allowing one to describe for instance the gauging of electromagnetism in our extension of QCD by the $S^2$ pions. 

\subsection{Non-invertible symmetry: a first look in the IR}

To be concrete, let us for now take $N=3$ flavour QCD, and consider gauging the vector-like $\U_Q \subset \SU(3)_{L+R}$ generated by 
\begin{equation}
    Q=\begin{pmatrix}
        2 & & \\
        & -1 & \\
        & & -1
    \end{pmatrix}   \, .
\end{equation}
Let $f_Q=da$ denote the corresponding abelian field strength.\footnote{Note that we have redefined $a$ with respect to previous Sections to here denote the abelian gauge field for electromagnetism (rather than the QCD gluon field as before). }
As alluded to above in \S \ref{sec:puzzle} (and used in~\cite{Davighi:2024zip}),
there is a term (amongst others) in the gauged mixed WZW action like
\begin{equation}
    S_{\text{WZW}} \supset 2\pi n_c \int_{\Sigma_4} \frac{\pi_0}{2\pi f_\pi} \frac{f_Q}{2\pi} \wedge \Vol\, .
\end{equation}
This arises because $\Tr[(t_L^3-t_R^3), Q] \neq 0$, where recall $t_L^3-t_R^3$ is the generator of the neutral pion shift symmetry, where we adopt the usual Gell-Mann basis for $\mathfrak{su}(3)$.

This term in the action now encodes not the 2-group global symmetry relation from before, but a genuine breaking of the global axial symmetry: doing a shift $\pi_0 \to \pi_0 + \alpha f_\pi$ gives
\begin{equation} \label{eq:NIS-shift}
    \delta_\alpha S_{\text{WZW}} = 2\pi n_c \int_{\Sigma_4} \frac{\alpha}{2\pi} \frac{f_Q}{2\pi} \wedge \Vol\, \, ,
\end{equation}
which can be non-zero for instance when evaluated on a spacetime manifold with topology $\Sigma_4 = S^2 \times S^2$.
This is analogous to the breaking of a global symmetry via an abelian ABJ anomaly --- or, in modern parlance, a non-invertible symmetry~\cite{Choi:2022jqy, Cordova:2022ieu} --- but with a mixed operator involving both $f_Q$ and the $S^2$ winding number appearing in the anomalous variation.

\subsection{Non-invertible symmetry from the UV anomaly polynomial}

This non-invertible symmetry structure can be traced up to our UV completion (\S \ref{sec:proper-completion}) via QCD coupled to SED. There, as we pass to the UV, the winding number on $S^2$ becomes identified with the field strength for the $\U$ gauge field $b$ from before (that couples to baryon number on the QCD side). In this UV theory, which now has two gauged $\U$ factors that we call $\U_Q$ and $\U_b$ in what we hope is an obvious notation, there is an abelian ABJ anomaly between the global axial current generating the pion shift and the two different gauged $\U$ groups. That is, a term in the anomaly polynomial $\propto F \wedge f_{Q} \wedge h$ is responsible for the anomalous shift (\ref{eq:NIS-shift}) in the UV theory, where $F$ is the background field strength for the axial current $\U_A$ under consideration, and $h=db$ still.

Let's see how this works more explicitly. The quantum numbers of the quarks under the various gauge groups and the $\U_A$ chiral global symmetry generated by $t_L^3 - t_R^3$ are given in Table \ref{tab:q-no-quarks-EM-gauged} below.
\begin{table}[h!]
  \centering
  \begin{tabular}{c|cccc}
    & $\SU(n_c)$ & $\U_{Q}$ & $\U_{\text{SED}}$ & $\U_{A}$\\
    \hline
    $\psi_1$ & ${\bf n_c}$ & $+2$ & $+1$ & $+1$\\
    $\psi_2$ & ${\bf n_c}$ & $-1$ & $+1$ & $-1$\\
    $\psi_3$ & ${\bf n_c}$ & $-1$ & $+1$ & $0$\\
    \hline
    $\tilde{\psi}_1$ & $\overline{\bf n_c}$ & $-2$ & $-1$ & $+1$\\
    $\tilde{\psi}_2$ & $\overline{\bf n_c}$ & $+1$ & $-1$ & $-1$\\
     $\tilde{\psi}_3$ & $\overline{\bf n_c}$ & $+1$ & $-1$ & $0$\\
  \end{tabular}
  \caption{Quantum numbers of the quark fields under various $\U$ symmetries relevant to the gauged version of our theory.}
  \label{tab:q-no-quarks-EM-gauged}
\end{table}
Note that both the gauged $\U$s are vector-like and so trivially free of gauge anomalies. But because of the chiral nature of the global symmetry $\U_{A}$, it is possible that there may be anomalies proportional to the background gauge field $F$. Indeed, turning on the background field $F$ for $\U_A$, the anomaly polynomial reads
\begin{equation} \label{eq:Phi6Faxial}
\Phi_6(F) = \frac{3 n_c}{(2\pi)^3} \left( F\wedge f_Q\wedge f_Q + 2 F\wedge f_Q\wedge h \right)\, .
\end{equation}
The first term on the right-hand-side encodes the usual ABJ anomaly between $\U_A$ and $\U_Q$ responsible for $\pi^0 \to \gamma \gamma$ decay in real-world QCD, while the second term encodes the new effect that is our main interest here. 

With recent advances in
our understanding of generalised symmetries, we know that a
non-invertible symmetry emerges from the $\U_A$ symmetry destroyed by this pair
of anomalies \cite{Choi:2022jqy, Cordova:2022ieu}. The symmetry defect will take the form
\begin{equation}
U_{\beta} = \exp \left( 2\pi i \beta\int_{M_3} \star J - i\frac{3n_c\beta}{2\pi}\int_{M_3}(a+2b)\wedge f \right), \quad\beta\in [0,1)\, ,
\end{equation}
with $\beta$ being rational.\footnote{For a slightly different take on the range of the transformation's parameter, see Refs. \cite{Karasik:2022kkq, GarciaEtxebarria:2022jky, Arbalestrier:2024oqg}.} The ill-defined CS terms are shorthands for
3d topological quantum field theories (TQFTs) localised on the defect submanifold $M_3$ which couple to the
$\U$ fields. For example, if we take $\beta = 1/6n_c K$, the symmetry defect becomes
\begin{equation}
    U_{1/6n_c K} = \exp \left( \frac{2\pi i }{6n_c K} \int_{M_3} \star J - \frac{i}{4\pi K} \int_{M_3} [(a+b)\wedge d(a+b) - b\wedge db]\right)\, .
\end{equation}
The second term in the exponential consists of two fractional CS theories, 
which can be properly defined with help from two auxiliary dynamical\footnote{This means $c_1$ 
and $c_2$ must be path-integrated implicitly in the definition of $U_{1/6n_c K}$. } $\U$ gauge fields 
$c_1$ and $c_2$, localised on $M_3$, via
\begin{equation}
    i\int_{M_3} \left(\frac{K}{4\pi} c_1 dc_1 - \frac{K}{4\pi} c_2 dc_2\right) + i\int_{M_3} \left(\frac{1}{2\pi} c_1 (da+db) - \frac{1}{2\pi} c_2 db\right)\,,
    \label{eq:fractional-Hall}
\end{equation}
The first term indicates that this TQFT is the $\U_K \times \U_{-K}$ CS theory, 
while the second term specifies the coupling between the auxiliary fields and the 4d $\U$ fields $a$ and $b$.
For other values of $\beta \in \mathbb{Q}$ that result in the CS coefficient being $p/K$ instead of 
the `unit fraction' $1/K$, Eq.~\eqref{eq:fractional-Hall} can be generalised in terms of a certain
minimal $\Z_K$ TQFT, usually denoted by $\mathcal{A}^{N,p}$ \cite{Choi:2022jqy}, which we will not elaborate on
further here. Interested readers are invited to see Refs. \cite{Gaiotto:2014kfa, Hsin:2018vcg} 
for detailed study of this particular TQFT.

\subsection{Full global structure: 2-group plus non-invertible symmetry}

One can consider turning on background gauge fields for other global symmetry currents that remain after gauging $\U_Q$.
To analyse the symmetry structure more generally, not just for the specific $\U_A$ subgroup in Table~\ref{tab:q-no-quarks-EM-gauged} that shifts the $\pi^0$, we first have to work out what the remaining flavour symmetry $G^{\prime}_{\text{flavour}}$ is after gauging. 

We claim that the faithful global symmetry becomes
\begin{equation} \label{eq:Gflavour}
G^{\prime}_{\text{flavour}}\,\cong\, \frac{\mathrm{U}(2) _L \times \mathrm{U}(2)_R}{\U}\, .
\end{equation}
To see this, note first that we can still treat
the left- and the right-handed quarks independently. Since the $\U$ EM
subgroup is diagonal, what happens to the left-handed component must
be the same as what happens to the right-handed component. The maximal
subgroup of $\SU(3)_L$ that commutes with the $\U$ subgroup of
$\SU(3)_L$ generated by $Q$ consists of matrices
of the form
\begin{equation}
 U = \begin{pmatrix}
     e^{i \phi} & 0\\
     0 & V
  \end{pmatrix}
\end{equation}
where $V$ is a $\mathrm{U}(2)$ matrix, so it can be written as
$V=e^{i \theta} \tilde{V}$ with $\tilde{V}\in \SU(2)$. Since the matrix
$U$ must still be unimodular, we need $\det U = 1$, or
$e^{i \phi + 2i \theta} = 1$. This means we need $U$ to be of the form
\begin{equation}
U =
\begin{pmatrix}
  e^{-2i \theta} & 0\\
  0 & e^{i \theta} \tilde{V}
\end{pmatrix}
\end{equation}
Matrices of this form form the group
\begin{equation}
    S \left[ \U \times \SU(2) \right]_L \cong \mathrm{U}(2)\, ,    
\end{equation}
where the isomorphism follows from the fact that the top-left entry is entirely fixed by the $\mathrm{U}(2)$ matrix in the bottom-right.
However, we should not be too quick
to conclude that the remaining global symmetry is
$\tilde{G}^{'} = \mathrm{U}(2)_L \times \mathrm{U}(2)_R
$. The catch comes from the fact that an element $u$
of the gauged $\U_Q \subset \SU(3)_L \times \SU(3)_R$ subgroup
takes the form
\begin{equation}
\left\{
 u = \begin{pmatrix}
    e^{2i \alpha} &&\\
    & e^{-i \alpha}&\\
    && e^{-i \alpha}
  \end{pmatrix}
  ,
  \begin{pmatrix}
    e^{-2i \alpha}&&\\
    & e^{i \alpha} &\\
    && e^{i \alpha}
  \end{pmatrix}
\right\}\, ,
\end{equation}
which is in the centre of $\tilde{G}^{\prime}$. Therefore, two elements of
$\tilde{G}^{\prime}$ related to each other by $u$ must be identified. This
reduces the full symmetry from $\tilde{G}^{\prime}$ down to the
$G^{\prime}_{\text{flavour}}$ group written in Eq. \eqref{eq:Gflavour}, as claimed.

From here we can determine the symmetry/anomaly structure of this theory, now turning on background gauge fields for a general global symmetry current (rather than just for the particular $\U_A$ choice, as in \eqref{eq:Phi6Faxial}). To do so, it is convenient to start from the anomaly polynomial $\Phi_6$ before gauging $\U_Q$, namely
\begin{equation}
    \Phi_6(F_L,F_R)=\frac{n_c}{3!} \frac{1}{(2\pi)^3} \left[ \Tr F_L^3 - \Tr F_R^3 \right]+ \frac{n_c}{2} \frac{h}{2\pi} \left[ \Tr \left( \frac{F_L}{2\pi} \right)^2 - \Tr \left( \frac{F_R}{2\pi} \right)^2 \right]\, , \end{equation}
and replace the non-abelian background gauge fields
$F_L$ and $F_R$ by the combinations
\begin{align}
    F_L &= f_Q Q + F^{\prime}_L\, , \\
    F_R &= f_Q Q + F^{\prime}_R\, ,
\end{align}
where $F^{\prime}_L$ and $F^{\prime}_R$ are the background fields for the global symmetry that remains after gauging $\U_Q$, which means they must commute with the gauge group, {\em i.e.} must satisfy
\begin{equation}
    [Q, F^{\prime}_L]= [Q, F^{\prime}_R]=0\, .
\end{equation}
After making this replacement, the anomaly
polynomial becomes
\begin{equation}\label{eq:Phi6FLRprime}
\Phi_6(F_L, F_R) = \Phi_6(F_L^{\prime}, F_R^{\prime}) + \frac{n_c}{(2\pi)^3} \left[ \frac{1}{2}f_Q^2 \Tr \left( Q^2(F^{\prime}_L-F^{\prime}_R) \right) + h f_Q \Tr \left( Q (F^{\prime}_L- F^{\prime}_R) \right)\right]\, ,
\end{equation}
where wedge products between the various 2-form field strengths are implicit.
Locally, we can write $F_L^{\prime}$ and $F_R^{\prime}$ as elements of the Lie
algebra of $S[\U \times \mathrm{U}(2)]_{L/R} \,\cong \,\mathrm{U}(2)_{L/R}$. Letting $\mathcal{F}_L \in \mathrm{U}(2)_L$, and
$\mathcal{F}_R\in \mathrm{U}(2)_R$, we can write
\begin{equation}
F^{\prime}_L =
\begin{pmatrix}
  -\Tr \mathcal{F}_L &\\
  & \mathcal{F}_L
\end{pmatrix}, \qquad F^{\prime}_R =
\begin{pmatrix}
  -\Tr \mathcal{F}_R &\\
  & \mathcal{F}_R
\end{pmatrix}\, .
\end{equation}
Substituting into \eqref{eq:Phi6FLRprime} and evaluating the traces, the anomaly polynomial
$\Phi_6$ reduces to
\begin{equation}
\Phi_6 = \Phi_6^{\prime} - \frac{3n_c}{(2\pi)^3}\left( \frac{f_Q^2}{2} + h f_Q\right) \left(\Tr \mathcal{F}_L - \Tr \mathcal{F}_R \right),
\end{equation}
where we use $\Phi^{\prime}_6 = \Phi_6(F^{\prime}_L, F^{\prime}_R)$ as a shorthand. 

\paragraph{Non-invertible symmetry part.}
The last term of the anomaly polynomial indicates that axial
generators in $\mathrm{U}(2)_{L/R}$ with non-vanishing trace, which can in general be expressed as a linear combination of the pion shift generator picked out in \eqref{eq:Phi6Faxial} plus a $t^8_L - t^8_R$ component (that also shifts the $\eta$ meson),
suffer a mixed abelian ABJ anomaly with $h \wedge f_Q$, becoming
non-invertible in the process~\cite{Choi:2022jqy, Cordova:2022ieu}. As described above, this is in addition to the usual contribution $\propto f_Q^2$ that is already there for pure QCD with gauged electromagnetism.

\paragraph{2-group symmetry part.}
We next examine the part $\Phi^{\prime}_6$ to show that there remains a subgroup of the flavour symmetry $G^\prime_{\mathrm{flavour}}$ that participates in a 2-group structure after our gauging of $\U_Q$, that links the 0-form flavour symmetry to the magnetic 1-form symmetry for $\U_b$. Using the fact that
$\mathcal{F}_L$ and $\mathcal{F}_R$ are $\mathrm{U}(2)$ field strengths, we can write them as 
\begin{equation}
\mathcal{F}_{L/R} = \frac{1}{2}\Tr{\mathcal{F}}_{L/R}\, \mathbf{1}_2 + \tilde{\mathcal{F}}_{L/R}
\end{equation}
where 
$\tilde{\mathcal{F}}_{L/R}$ are $\su(2)$-valued. The part of the anomaly polynomial denoted
$\Phi^{\prime}_6$ then reads
\begin{align}
  \Phi^{\prime}_6 =\, &\frac{n_c}{3!}\frac{\Tr{\mathcal{F}}_L}{(2\pi)^3} \left[ -\frac{3}{4}(\Tr{\mathcal{F}}_L)^2 + \frac{3}{2} \Tr (\tilde{\mathcal{F}}_L^2) \right] \\
  + &\frac{n_c}{2}\frac{h}{(2\pi)^3} \left[ \frac{3}{2} (\Tr{\mathcal{F}}_L)^2 + \Tr (\tilde{\mathcal{F}}_L^2) \right] - \left( L \leftrightarrow R \right)\, .
\end{align}
Let us digest the various terms appearing here:
\begin{itemize}
    \item The top line in this expansion of $\Phi^{\prime}_6$ captures the ordinary 't Hooft anomaly in $ \mathrm{U}(2)_{L/R}$. 
    \item The second line partly tells us that there is still the 2-group structure between the 1-form symmetry and the $\SU(2)_{L/R}$ part of the flavour symmetry with the same Postnikov classes as before. Focussing on the non-abelian part $\propto h \wedge \Tr (\tilde{\mathcal{F}}_L^2)$, the interpretation of this 2-group global symmetry is the same as before. For instance, one can consider the axial transformation generated by \begin{equation}
    X_L = -X_R = \begin{pmatrix}
        0& & \\
        & 1 & \\
        & & -1
    \end{pmatrix}   \, ,
    \end{equation}
    which does not suffer from any ABJ-like anomaly because 
    \begin{equation}
    \Tr (X Q^2)=\Tr (X Q)=0    
    \end{equation}
    and so defines a proper invertible symmetry. This $X$ (together with a whole $\mathfrak{su}(2)$ subalgebra in the lower-right block) participates in a 2-group current algebra with $h$, much like before but reduced from $\SU(3)$s to $\SU(2)$s. 
    \item The abelian part $\propto h \wedge (\Tr {\mathcal{F}}_L^2)$, however, introduces a qualitatively new kind of symmetry structure:
    there is a non-trivial 2-group relation between the magnetic 1-form symmetry and the abelian part of the remaining flavour symmetry which is itself non-invertible, as we previously explained.
\end{itemize}
To our knowledge, this kind of global symmetry (namely one which, in the old terminology, participates in both ABJ anomalies and operator-valued mixed anomalies with gauged currents) has not been studied before, and will be explored in future work.

\section{Conclusions and Outlook}

Topological WZW terms play a fundamental role in low-energy theories of pions that arise from confining gauge theories: they are needed to match chiral anomalies. Our starting point in this paper is a curious example of a pion EFT, on a coset $\SU(N) \times S^2$, that features a WZW term {\em not} related to any underlying chiral anomaly. While this term defines a perfectly sensible-looking EFT, it is not {\em a priori} clear how this might arise from a microscopic theory, and why its coefficient should end up quantized if not mandated by anomaly matching.

We find that this WZW term encodes not an anomaly, but a generalised 2-group global symmetry structure (that mixes ordinary flavour symmetries with a 1-form symmetry). Like the anomaly, this 2-group structure is rigid, as manifest at low-energies in the quantization condition for the WZW term, and so can be used to check the consistency of possible UV completions. Through this notion of `symmetry matching' (as opposed to anomaly matching), we rule out na\"ive QCD-like completions of this pion EFT, and instead propose a weakly-coupled completion involving a QCD sector and a $\mathbb{C}P^1$ sector, coupled weakly together in a very particular way by an abelian gauge field that gets Higgsed along the RG flow. Strikingly, by integrating out the weakly coupled gauge boson at tree-level one ends up with a quantized topological term -- which moreover implies the tree-level matching is exact. Even though loop corrections are expected to vanish, a `global' form of EFT matching (topologically-speaking) is nonetheless needed to match precisely, which is conveniently handled by using homogeneous coordinates on $\mathbb{C}P^1$. This RG flow is a highly non-generic one, and occurs because the abelian gauge field couples to topologically non-trivial currents in both the QCD and $\mathbb{C}P^1$ sectors.

We discussed several variations of the setup, for instance, several alternatives to the $\mathbb{C}P^1$ model on the scalar side. We furthermore examined the important scenario in which an anomaly-free subgroup of the QCD flavour symmetry (such as QED) is gauged. This gives rise to a more complicated generalised symmetry structure involving both 2-group and non-invertible symmetry. This also points to a new kind of symmetry structure, that we wish to study in future work, since the theory necessarily yields global symmetry currents $F$ that participate simultaneously in mixed anomalies of both `$Fff$' and `$FFf$' type. In addition, we wish to study this phenomenon of symmetry matching by WZW terms in a more general context, including discrete and/or non-perturbative examples, and theories outside 3+1 dimensions. 

Lastly, turning to phenomenology, we aim to apply these ideas to investigate UV completions of the topological portal to dark matter proposed in~\cite{Davighi:2024zip}. With an explicit (and weakly-coupled) model in hand, one can compute the full set of phenomenological predictions in both collider and cosmological observables, to characterise the viable parameter space and the best probes of this scenario.

\section*{Acknowledgement}

We thank Philip Boyle-Smith, Hitoshi Murayama, Yuji Tachikawa, David Tong, and Ethan Torres for stimulating discussions. We are further grateful to Yuji Tachikawa for detailed comments on the draft.
JD thanks Admir Greljo and Nud\v zeim Selimovi\'c for collaborating on the related project~\cite{Davighi:2024zip}.
NL is supported by the STFC consolidated grant in Particles, Strings and Cosmology number ST/T000708/1 and the Royal Society of London. 
We thank King's College London for hosting us for 1 productive week spent working on this project, and lastly we thank the Watch House Caf\'e, Somerset House, for providing us with excellent coffee in this time. 

\appendix
\section{Computation of \texorpdfstring{$\tilde{\Omega}^{\Spin}_4 \left(\SU(N) \times S^2\right)$}{the reduced 4th spin bordism group of SU(N)×S²}}
\label{app:bordism-comp}

In this Appendix, we present our computation of the reduced spin
bordism groups of $\SU(N)\times S^2$ up to degree $4$, using the Adams
spectral sequence (ASS) \cite{Adams:1958}. In particular, we will show
that
\begin{equation} \label{eq:appOm4}
\tilde{\Omega}^{\Spin}_4 \left( \SU(N)\times S^2 \right) \cong \Z_2\,.
\end{equation}
A readable practice guide on how to use the ASS to compute bordism
groups can be found in {\it e.g.} Refs. \cite{beaudry2018guide, Campbell:2017khc} 
(see also Appendix A1 of Ref. \cite{Davighi:2023mzg} for a brief summary of the general method).

The ASS relevant for us is a cohomological spectral sequence, consisting of a sequence of bi-graded abelian
groups $E_r^{s,t}$, $r=2,3,4,\ldots$ with gradings $s,t \geq 0$, together with
differentials $d_r: E_r^{s,t}\to E^{s+r,t-r+1}_r$ forming layers of
cochain complexes. We refer to the set of entries for a fixed value of the index $r$ as a `page' of the
sequence. An element $E^{s,t}_r$ of a page is determined from the
previous page by taking the homology with respect to the differentials:
\begin{equation}
E^{s,t}_{r+1} = \frac{\ker \left( d_r:E^{s,t}_{r} \to E^{s+r, t-r+1}_r \right)}{\mathrm{im}\left( d_r: E^{s-r, t+r-1}_r \to E^{s,t}_r \right)}
\end{equation}
For our purpose, the initial $E_2$ page of the ASS, and what it
converges to, is given by
\begin{equation}
E^{s,t}_2 = \ext^{s,t}_{\mathcal{A}(1)} \left( \tilde{H}^{\bullet}(\SU(N)\times S^2;\Z_2), \Z_2 \right) \Longrightarrow \left( \tilde{\Omega}^{\Spin}_{t-s} \left( \SU(N) \times S^2 \right) \right)^{\wedge}_2\,,
\end{equation}
where $\mathcal{A}(1)$ is the Steenrod subalgebra spanned by the Steenrod
squares operations $\left\{1, \sq^1, \sq^2\right\}$, and $(\cdot)^{\wedge}_2$ denotes the
2-completion. Strictly speaking, this ASS will only give us
information about the free and the 2-torsion parts of the bordism
groups. In our case, however, we can check by other means, such as the
Atiyah--Hirzebruch spectral sequence (AHSS), that there are no other torsions
present.\footnote{We find the AHSS on its own, however, is not sufficient to deduce the bordism group \eqref{eq:appOm4} due to unknown differentials. }
We therefore obtain complete information regarding the bordism
groups we want to compute using this method -- provided we can solve the
spectral sequence.

\begin{figure}[h!]
  \centering
  \includegraphics[scale=0.85]{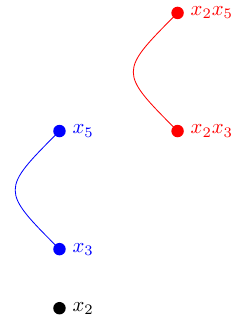}
  \caption{$\tilde{H}^{\bullet}(\SU(N) \times S^2)$ as an $\mathcal{A}(1)$-module.}
  \label{fig:HSUN-S2-A1}
\end{figure}

To start, we need to know the mod 2 cohomology ring of
$\SU(N)\times S^2$ as an $\mathcal{A}(1)$-module. The mod 2 cohomology ring of
$\SU(N)$ is given by \cite{Borel-Serre:1953a}
\begin{equation}
H^{\bullet}(\SU(N);\Z_2) \cong \bigwedge\nolimits_{\Z_2} \left[ x_3,x_5,\ldots, x_{2N-1} \right]\,,
\end{equation}
that is, it is the exterior algebra on generators $x_3$, $x_5$,
$\ldots$, $x_{2N-1}$ with integer mod 2 coefficients, where $x_i$ are
generators in degree $i$. The $\mathcal{A}(1)$-module structure is given by the
action of the Steenrod squares on the generators:
\begin{equation}
\sq^1 x_i = 0, \qquad \sq^2 x_{2j-1} = \binom{j-1}{i} x_{2i+2j-1}\,.
\end{equation}
Similarly, we can write the mod 2 cohomology ring of
$S^2$ as
\begin{equation}
H^{\bullet}(S^2;\Z_2) \cong \bigwedge\nolimits_{\Z_2} \left[ x_2 \right]\,,
\end{equation}
with $\sq^1x_2 = \sq^2 x_2 = 0$ on dimensional grounds. Applying
K\"unneth's theorem, we obtain
\begin{equation}
H^{\bullet}(\SU(N)\times S^2;\Z_2) \cong \bigwedge\nolimits_{\Z_2}\left[ x_2, x_3,x_5,\ldots, x_{2N-1} \right]\,.
\end{equation}
The graphical representation of its reduced version ({\em i.e.} that obtained by ignoring the
degree 0 generator) as an $\mathcal{A}(1)$-module is shown in
Fig. \ref{fig:HSUN-S2-A1}, up to degree $5$. Straight lines (of which there happen to be none) represent the action of
$\sq^1$, while the curved lines represent the action of $\sq^2$.

\begin{figure}[h!]
  \centering
  \includegraphics[scale=0.8]{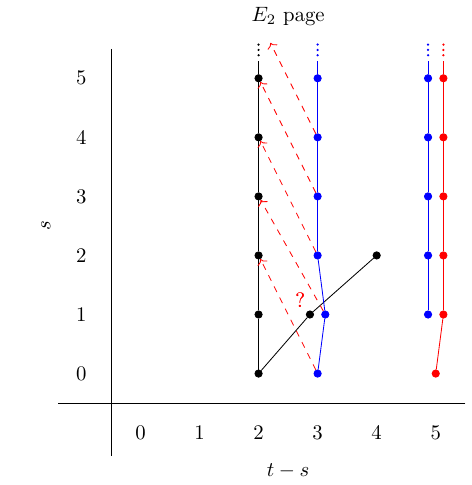}
  \caption{The Adams chart for $\ext^{s,t}_{\mathcal{A}(1)}\left( H^{\bullet}(\SU(N)\times S^2),\Z_2 \right)$.}
  \label{fig:Adams-chart-HSUN-S2}
\end{figure}

We are now ready to compute the $E_2$ page of the spectral sequence,
which is given by
$E^{s,t}_2 = \ext^{s,t}_{\mathcal{A}(1)}\left( H^{\bullet}(\SU(N)\times S^2),\Z_2 \right)$. As this is
a direct sum of the functor
$\ext^{s,t}_{\mathcal{A}(1)}(\text{--},\Z_2)$ applied on each connected component
in Fig. \ref{fig:HSUN-S2-A1}, which are known \cite{beaudry2018guide},
we simply stitch them together and present the result graphically in
Fig. \ref{fig:Adams-chart-HSUN-S2} in what is known as the `Adams
chart' for $\ext^{s,t}_{\mathcal{A}(1)}\left( H^{\bullet}(\SU(N)\times S^2),\Z_2 \right)$.

We next solve the spectral sequence by `turning the pages', taking
successive homology with respect to the differentials to go from one
$E_r$ page to the next. This is done until no elements in the range we
are interested in change any more, at which point we say that the
sequence stabilises. On the $E_2$ page, the only possible non-trivial
differentials $d_2: E^{s,t}_2 \to E^{s+2,t+1}_2$ in the range of $t-s$
that we are interested in are from the column $t-s = 3$ to the column
$t-s = 2$, as indicated in Fig. \ref{fig:Adams-chart-HSUN-S2}. By a
comparison with the AHSS, we can easily determine that these
differentials, as well as similar differentials on subsequent pages, are trivial. The spectral sequence in the relevant range
$t-s\leq 4$ therefore stabilises already on this page. We can then read
off the reduced bordism groups directly from the Adams chart, shown in
Table \ref{tab:red-bord-SUN-S2} below. In particular, the reduced spin
bordism group of $\SU(N)\times S^2$ in degree $4$ is $\Z_2$ as claimed.
\begin{table}[h!]
  \centering
  {\renewcommand{\arraystretch}{1.2}
  \begin{tabular}{|c|ccccc|}
    \hline
    $i$ & $0$ & $1$ & $2$ & $3$ & $4$\\
    \hline
    $\tilde{\Omega}^{\Spin}_i(\SU(N)\times S^2)$ & $0$ & $0$ & $\Z$ & $\Z \oplus \Z_2$ & $\Z_2$\\
    \hline
  \end{tabular}
  }
  \caption{The reduced spin bordism groups of $\SU(N) \times S^2$ up to degree $4$.}
  \label{tab:red-bord-SUN-S2}
\end{table}

\bibliographystyle{JHEP}
\bibliography{references}

\providecommand{\href}[2]{#2}\begingroup\raggedright\begin{thebibliography}{10}

\bibitem{Nambu:1960xd}
Y.~Nambu, \emph{{Axial vector current conservation in weak interactions}},
  \href{https://doi.org/10.1103/PhysRevLett.4.380}{\emph{Phys. Rev. Lett.}
  {\bfseries 4} (1960) 380--382}.

\bibitem{Gell-Mann:1960mvl}
M.~Gell-Mann and M.~Levy, \emph{{The axial vector current in beta decay}},
  \href{https://doi.org/10.1007/BF02859738}{\emph{Nuovo Cim.} {\bfseries 16}
  (1960) 705}.

\bibitem{Adler:1969gk}
S.~L. Adler, \emph{{Axial vector vertex in spinor electrodynamics}},
  \href{https://doi.org/10.1103/PhysRev.177.2426}{\emph{Phys. Rev.} {\bfseries
  177} (1969) 2426--2438}.

\bibitem{Bell:1969ts}
J.~S. Bell and R.~Jackiw, \emph{{A PCAC puzzle: $\pi^0 \to \gamma \gamma$ in
  the $\sigma$ model}}, \href{https://doi.org/10.1007/BF02823296}{\emph{Nuovo
  Cim. A} {\bfseries 60} (1969) 47--61}.

\bibitem{Wess:1971yu}
J.~Wess and B.~Zumino, \emph{{Consequences of anomalous Ward identities}},
  \href{https://doi.org/10.1016/0370-2693(71)90582-X}{\emph{Phys. Lett. B}
  {\bfseries 37} (1971) 95--97}.

\bibitem{tHooft:1976rip}
G.~'t~Hooft, \emph{{Symmetry Breaking Through Bell-Jackiw Anomalies}},
  \href{https://doi.org/10.1103/PhysRevLett.37.8}{\emph{Phys. Rev. Lett.}
  {\bfseries 37} (1976) 8--11}.

\bibitem{Witten:1979vv}
E.~Witten, \emph{{Current Algebra Theorems for the U(1) Goldstone Boson}},
  \href{https://doi.org/10.1016/0550-3213(79)90031-2}{\emph{Nucl. Phys. B}
  {\bfseries 156} (1979) 269--283}.

\bibitem{Veneziano:1979ec}
G.~Veneziano, \emph{{U(1) Without Instantons}},
  \href{https://doi.org/10.1016/0550-3213(79)90332-8}{\emph{Nucl. Phys. B}
  {\bfseries 159} (1979) 213--224}.

\bibitem{Witten:1983tw}
E.~Witten, \emph{{Global Aspects of Current Algebra}},
  \href{https://doi.org/10.1016/0550-3213(83)90063-9}{\emph{Nucl. Phys. B}
  {\bfseries 223} (1983) 422--432}.

\bibitem{Gaiotto:2014kfa}
D.~Gaiotto, A.~Kapustin, N.~Seiberg and B.~Willett, \emph{{Generalized Global
  Symmetries}}, \href{https://doi.org/10.1007/JHEP02(2015)172}{\emph{JHEP}
  {\bfseries 02} (2015) 172},
  [\href{https://arxiv.org/abs/1412.5148}{{\ttfamily 1412.5148}}].

\bibitem{Gaiotto:2017yup}
D.~Gaiotto, A.~Kapustin, Z.~Komargodski and N.~Seiberg, \emph{{Theta, Time
  Reversal, and Temperature}},
  \href{https://doi.org/10.1007/JHEP05(2017)091}{\emph{JHEP} {\bfseries 05}
  (2017) 091}, [\href{https://arxiv.org/abs/1703.00501}{{\ttfamily
  1703.00501}}].

\bibitem{Kapustin:2013uxa}
A.~Kapustin and R.~Thorngren, \emph{{Higher Symmetry and Gapped Phases of Gauge
  Theories}}, \href{https://doi.org/10.1007/978-3-319-59939-7_5}{\emph{Prog.
  Math.} {\bfseries 324} (2017) 177--202},
  [\href{https://arxiv.org/abs/1309.4721}{{\ttfamily 1309.4721}}].

\bibitem{Baez:2004in}
J.~Baez and U.~Schreiber, \emph{{Higher gauge theory: 2-connections on
  2-bundles}},  \href{https://arxiv.org/abs/hep-th/0412325}{{\ttfamily
  hep-th/0412325}}.

\bibitem{Baez:2005qu}
J.~C. Baez and U.~Schreiber, \emph{{Higher gauge theory}},
  \href{https://arxiv.org/abs/math/0511710}{{\ttfamily math/0511710}}.

\bibitem{Baez:2010ya}
J.~C. Baez and J.~Huerta, \emph{{An Invitation to Higher Gauge Theory}},
  \href{https://doi.org/10.1007/s10714-010-1070-9}{\emph{Gen. Rel. Grav.}
  {\bfseries 43} (2011) 2335--2392},
  [\href{https://arxiv.org/abs/1003.4485}{{\ttfamily 1003.4485}}].

\bibitem{Sharpe:2015mja}
E.~Sharpe, \emph{{Notes on generalized global symmetries in QFT}},
  \href{https://doi.org/10.1002/prop.201500048}{\emph{Fortsch. Phys.}
  {\bfseries 63} (2015) 659--682},
  [\href{https://arxiv.org/abs/1508.04770}{{\ttfamily 1508.04770}}].

\bibitem{Cordova:2020tij}
C.~Cordova, T.~T. Dumitrescu and K.~Intriligator, \emph{{2-Group Global
  Symmetries and Anomalies in Six-Dimensional Quantum Field Theories}},
  \href{https://doi.org/10.1007/JHEP04(2021)252}{\emph{JHEP} {\bfseries 04}
  (2021) 252}, [\href{https://arxiv.org/abs/2009.00138}{{\ttfamily
  2009.00138}}].

\bibitem{Davighi:2024zip}
J.~Davighi, A.~Greljo and N.~Selimovic, \emph{{Topological Portal to the Dark
  Sector}},  \href{https://arxiv.org/abs/2401.09528}{{\ttfamily 2401.09528}}.

\bibitem{Iqbal:2020lrt}
N.~Iqbal and N.~Poovuttikul, \emph{{2-group global symmetries, hydrodynamics
  and holography}},
  \href{https://doi.org/10.21468/SciPostPhys.15.2.063}{\emph{SciPost Phys.}
  {\bfseries 15} (2023) 063},
  [\href{https://arxiv.org/abs/2010.00320}{{\ttfamily 2010.00320}}].

\bibitem{DeWolfe:2020uzb}
O.~DeWolfe and K.~Higginbotham, \emph{{Generalized symmetries and 2-groups via
  electromagnetic duality in $AdS/CFT$}},
  \href{https://doi.org/10.1103/PhysRevD.103.026011}{\emph{Phys. Rev. D}
  {\bfseries 103} (2021) 026011},
  [\href{https://arxiv.org/abs/2010.06594}{{\ttfamily 2010.06594}}].

\bibitem{Lee:2021crt}
Y.~Lee, K.~Ohmori and Y.~Tachikawa, \emph{{Matching higher symmetries across
  Intriligator-Seiberg duality}},
  \href{https://doi.org/10.1007/JHEP10(2021)114}{\emph{JHEP} {\bfseries 10}
  (2021) 114}, [\href{https://arxiv.org/abs/2108.05369}{{\ttfamily
  2108.05369}}].

\bibitem{Skyrme:1961vq}
T.~H.~R. Skyrme, \emph{{A Nonlinear field theory}},
  \href{https://doi.org/10.1098/rspa.1961.0018}{\emph{Proc. Roy. Soc. Lond. A}
  {\bfseries 260} (1961) 127--138}.

\bibitem{Balachandran:1982dw}
A.~P. Balachandran, V.~P. Nair, S.~G. Rajeev and A.~Stern, \emph{{Exotic Levels
  from Topology in the QCD Effective Lagrangian}},
  \href{https://doi.org/10.1103/PhysRevLett.49.1124}{\emph{Phys. Rev. Lett.}
  {\bfseries 49} (1982) 1124}.

\bibitem{Witten:1983tx}
E.~Witten, \emph{{Current Algebra, Baryons, and Quark Confinement}},
  \href{https://doi.org/10.1016/0550-3213(83)90064-0}{\emph{Nucl. Phys. B}
  {\bfseries 223} (1983) 433--444}.

\bibitem{Tucker-Smith:2001myb}
D.~Tucker-Smith and N.~Weiner, \emph{{Inelastic dark matter}},
  \href{https://doi.org/10.1103/PhysRevD.64.043502}{\emph{Phys. Rev. D}
  {\bfseries 64} (2001) 043502},
  [\href{https://arxiv.org/abs/hep-ph/0101138}{{\ttfamily hep-ph/0101138}}].

\bibitem{Freed:2006mx}
D.~S. Freed, \emph{{Pions and Generalized Cohomology}}, {\emph{J. Diff. Geom.}
  {\bfseries 80} (2008) 45--77},
  [\href{https://arxiv.org/abs/hep-th/0607134}{{\ttfamily hep-th/0607134}}].

\bibitem{Davighi:2018inx}
J.~Davighi and B.~Gripaios, \emph{{Homological classification of topological
  terms in sigma models on homogeneous spaces}},
  \href{https://doi.org/10.1007/JHEP09(2018)155}{\emph{JHEP} {\bfseries 09}
  (2018) 155}, [\href{https://arxiv.org/abs/1803.07585}{{\ttfamily
  1803.07585}}].

\bibitem{Lee:2020ojw}
Y.~Lee, K.~Ohmori and Y.~Tachikawa, \emph{{Revisiting Wess-Zumino-Witten
  terms}}, \href{https://doi.org/10.21468/SciPostPhys.10.3.061}{\emph{SciPost
  Phys.} {\bfseries 10} (2021) 061},
  [\href{https://arxiv.org/abs/2009.00033}{{\ttfamily 2009.00033}}].

\bibitem{Yonekura:2020upo}
K.~Yonekura, \emph{{General anomaly matching by Goldstone bosons}},
  \href{https://doi.org/10.1007/JHEP03(2021)057}{\emph{JHEP} {\bfseries 03}
  (2021) 057}, [\href{https://arxiv.org/abs/2009.04692}{{\ttfamily
  2009.04692}}].

\bibitem{Adams:1958}
J.~F. Adams, \emph{{On the structure and applications of the Steenrod
  algebra}}, \href{https://doi.org/10.1007/BF02564578}{\emph{Commentarii
  Mathematici Helvetici} {\bfseries 32} (1958) 180--214}.

\bibitem{Witten:1982fp}
E.~Witten, \emph{{An $SU(2)$ Anomaly}},
  \href{https://doi.org/10.1016/0370-2693(82)90728-6}{\emph{Phys. Lett.}
  {\bfseries B117} (1982) 324--328}.

\bibitem{bott1958space}
R.~Bott, \emph{{The space of loops on a Lie group}}, {\emph{Michigan
  Mathematical Journal} {\bfseries 5} (1958) 35--61}.

\bibitem{atiyah1968indexI}
M.~F. Atiyah and I.~M. Singer, \emph{{The index of elliptic operators: I}},
  {\emph{Annals of mathematics} {\bfseries 87} (1968) 484--530}.

\bibitem{atiyah1968indexII}
M.~F. Atiyah and I.~M. Singer, \emph{{The index of elliptic operators: III}},
  {\emph{Annals of mathematics} {\bfseries 87} (1968) 546--604}.

\bibitem{atiyah1968indexIII}
M.~F. Atiyah and G.~B. Segal, \emph{{The index of elliptic operators: II}},
  {\emph{Annals of Mathematics} {\bfseries 87} (1968) 531--545}.

\bibitem{tHooft:1979rat}
G.~'t~Hooft, \emph{{Naturalness, chiral symmetry, and spontaneous chiral
  symmetry breaking}},
  \href{https://doi.org/10.1007/978-1-4684-7571-5_9}{\emph{NATO Sci. Ser. B}
  {\bfseries 59} (1980) 135--157}.

\bibitem{Davighi:2023luh}
J.~Davighi, N.~Lohitsiri and A.~Debray, \emph{{Toric 2-group anomalies via
  cobordism}}, \href{https://doi.org/10.1007/JHEP07(2023)019}{\emph{JHEP}
  {\bfseries 07} (2023) 019},
  [\href{https://arxiv.org/abs/2302.12853}{{\ttfamily 2302.12853}}].

\bibitem{Cordova:2018cvg}
C.~C\'ordova, T.~T. Dumitrescu and K.~Intriligator, \emph{{Exploring 2-Group
  Global Symmetries}},
  \href{https://doi.org/10.1007/JHEP02(2019)184}{\emph{JHEP} {\bfseries 02}
  (2019) 184}, [\href{https://arxiv.org/abs/1802.04790}{{\ttfamily
  1802.04790}}].

\bibitem{Cordova:2022qtz}
C.~Cordova and S.~Koren, \emph{{Higher Flavor Symmetries in the Standard
  Model}}, \href{https://doi.org/10.1002/andp.202300031}{\emph{Annalen Phys.}
  {\bfseries 535} (2023) 2300031},
  [\href{https://arxiv.org/abs/2212.13193}{{\ttfamily 2212.13193}}].

\bibitem{Brennan:2020ehu}
T.~D. Brennan and C.~Cordova, \emph{{Axions, higher-groups, and emergent
  symmetry}}, \href{https://doi.org/10.1007/JHEP02(2022)145}{\emph{JHEP}
  {\bfseries 02} (2022) 145},
  [\href{https://arxiv.org/abs/2011.09600}{{\ttfamily 2011.09600}}].

\bibitem{Choi:2022fgx}
Y.~Choi, H.~T. Lam and S.-H. Shao, \emph{{Non-invertible Gauss law and
  axions}}, \href{https://doi.org/10.1007/JHEP09(2023)067}{\emph{JHEP}
  {\bfseries 09} (2023) 067},
  [\href{https://arxiv.org/abs/2212.04499}{{\ttfamily 2212.04499}}].

\bibitem{alexander1985differential}
J.~Cheeger and J.~Simons, \emph{Differential characters and geometric
  invariants}.
\newblock Springer, 1985.

\bibitem{bar2014differential}
C.~B{\"a}r and C.~Becker, \emph{Differential characters}, vol.~2112.
\newblock Springer, 2014.

\bibitem{Davighi:2020vcm}
J.~Davighi, B.~Gripaios and O.~Randal-Williams, \emph{{Differential cohomology
  and topological actions in physics}},
  \href{https://arxiv.org/abs/2011.05768}{{\ttfamily 2011.05768}}.

\bibitem{Wu:1976qk}
T.~T. Wu and C.~N. Yang, \emph{{Dirac's Monopole Without Strings: Classical
  Lagrangian Theory}},
  \href{https://doi.org/10.1103/PhysRevD.14.437}{\emph{Phys. Rev. D} {\bfseries
  14} (1976) 437--445}.

\bibitem{Alvarez:1984es}
O.~Alvarez, \emph{{Topological Quantization and Cohomology}},
  \href{https://doi.org/10.1007/BF01212452}{\emph{Commun. Math. Phys.}
  {\bfseries 100} (1985) 279}.

\bibitem{Bott-Tu:1982}
R.~Bott and L.~W. Tu, \emph{Differential forms in algebraic topology}, vol.~82
  of \emph{Graduate Texts in Mathematics}.
\newblock Springer-Verlag, New York, 1982.

\bibitem{tong2018gauge}
D.~Tong, \emph{Gauge theory}, {\emph{Lecture notes, DAMTP Cambridge} {\bfseries
  10} (2018) 8}.

\bibitem{DHoker:1994rdl}
E.~D'Hoker and S.~Weinberg, \emph{{General effective actions}},
  \href{https://doi.org/10.1103/PhysRevD.50.R6050}{\emph{Phys. Rev. D}
  {\bfseries 50} (1994) R6050--R6053},
  [\href{https://arxiv.org/abs/hep-ph/9409402}{{\ttfamily hep-ph/9409402}}].

\bibitem{Manton:1985jm}
N.~S. Manton, \emph{{The Schwinger Model and Its Axial Anomaly}},
  \href{https://doi.org/10.1016/0003-4916(85)90199-X}{\emph{Annals Phys.}
  {\bfseries 159} (1985) 220--251}.

\bibitem{Davighi:2018xwn}
J.~Davighi and B.~Gripaios, \emph{{Topological terms in Composite Higgs
  Models}}, \href{https://doi.org/10.1007/JHEP11(2018)169}{\emph{JHEP}
  {\bfseries 11} (2018) 169},
  [\href{https://arxiv.org/abs/1808.04154}{{\ttfamily 1808.04154}}].

\bibitem{Gripaios:2016mmi}
B.~Gripaios, M.~Nardecchia and T.~You, \emph{{On the Structure of Anomalous
  Composite Higgs Models}},
  \href{https://doi.org/10.1140/epjc/s10052-017-4603-5}{\emph{Eur. Phys. J. C}
  {\bfseries 77} (2017) 28},
  [\href{https://arxiv.org/abs/1605.09647}{{\ttfamily 1605.09647}}].

\bibitem{moroianu2007lectures}
A.~Moroianu, \emph{Lectures on K{\"a}hler geometry}, vol.~69.
\newblock Cambridge University Press, 2007.

\bibitem{Tachikawa:2017gyf}
Y.~Tachikawa, \emph{{On gauging finite subgroups}},
  \href{https://doi.org/10.21468/SciPostPhys.8.1.015}{\emph{SciPost Phys.}
  {\bfseries 8} (2020) 015},
  [\href{https://arxiv.org/abs/1712.09542}{{\ttfamily 1712.09542}}].

\bibitem{Choi:2022jqy}
Y.~Choi, H.~T. Lam and S.-H. Shao, \emph{{Noninvertible Global Symmetries in
  the Standard Model}},
  \href{https://doi.org/10.1103/PhysRevLett.129.161601}{\emph{Phys. Rev. Lett.}
  {\bfseries 129} (2022) 161601},
  [\href{https://arxiv.org/abs/2205.05086}{{\ttfamily 2205.05086}}].

\bibitem{Cordova:2022ieu}
C.~Cordova and K.~Ohmori, \emph{{Noninvertible Chiral Symmetry and Exponential
  Hierarchies}}, \href{https://doi.org/10.1103/PhysRevX.13.011034}{\emph{Phys.
  Rev. X} {\bfseries 13} (2023) 011034},
  [\href{https://arxiv.org/abs/2205.06243}{{\ttfamily 2205.06243}}].

\bibitem{Karasik:2022kkq}
A.~Karasik, \emph{{On anomalies and gauging of U(1) non-invertible symmetries
  in 4d QED}},
  \href{https://doi.org/10.21468/SciPostPhys.15.1.002}{\emph{SciPost Phys.}
  {\bfseries 15} (2023) 002},
  [\href{https://arxiv.org/abs/2211.05802}{{\ttfamily 2211.05802}}].

\bibitem{GarciaEtxebarria:2022jky}
I.~n. Garc\'\i{}a~Etxebarria and N.~Iqbal, \emph{{A Goldstone theorem for
  continuous non-invertible symmetries}},
  \href{https://doi.org/10.1007/JHEP09(2023)145}{\emph{JHEP} {\bfseries 09}
  (2023) 145}, [\href{https://arxiv.org/abs/2211.09570}{{\ttfamily
  2211.09570}}].

\bibitem{Arbalestrier:2024oqg}
A.~Arbalestrier, R.~Argurio and L.~Tizzano, \emph{{The Non-Invertible Axial
  Symmetry in QED Comes Full Circle}},
  \href{https://arxiv.org/abs/2405.06596}{{\ttfamily 2405.06596}}.

\bibitem{Hsin:2018vcg}
P.-S. Hsin, H.~T. Lam and N.~Seiberg, \emph{{Comments on One-Form Global
  Symmetries and Their Gauging in 3d and 4d}},
  \href{https://doi.org/10.21468/SciPostPhys.6.3.039}{\emph{SciPost Phys.}
  {\bfseries 6} (2019) 039},
  [\href{https://arxiv.org/abs/1812.04716}{{\ttfamily 1812.04716}}].

\bibitem{beaudry2018guide}
A.~Beaudry and J.~A. Campbell, \emph{A guide for computing stable homotopy
  groups},  \href{https://arxiv.org/abs/1801.07530}{{\ttfamily 1801.07530}}.

\bibitem{Campbell:2017khc}
J.~A. Campbell, \emph{{Homotopy Theoretic Classification of Symmetry Protected
  Phases}},  \href{https://arxiv.org/abs/1708.04264}{{\ttfamily 1708.04264}}.

\bibitem{Davighi:2023mzg}
J.~Davighi, N.~Lohitsiri and N.~Poovuttikul, \emph{{A non-perturbative mixed
  anomaly and fractional hydrodynamic transport}},
  \href{https://doi.org/10.1007/JHEP03(2024)119}{\emph{JHEP} {\bfseries 03}
  (2024) 119}, [\href{https://arxiv.org/abs/2311.18023}{{\ttfamily
  2311.18023}}].

\bibitem{Borel-Serre:1953a}
A.~Borel and J.-P. Serre, \emph{Groupes de lie et puissances réduites de
  steenrod}, {\emph{American Journal of Mathematics} {\bfseries 75} (1953)
  409--448}.

\end{thebibliography}\endgroup
\end{document}